\documentclass[11pt,preprint]{aastex}
\newcommand{\lsim}{\lower.5ex\hbox{$\; \buildrel < \over \sim \;$}}
\newcommand{\gsim}{\lower.5ex\hbox{$\; \buildrel > \over \sim \;$}}

\def \bea{\begin{eqnarray}}

\def \beq{\begin{equation}}

\def \eea{\end{eqnarray}}
\def \eeq{\end{equation}}

\def \U1Y{U(1)$_Y$}

\begin{document}
\slugcomment{Invited revew published in PASP 120, 235-265 (March 2008 issue)}

\title{The Beginning and Evolution of the Universe}

\author{Bharat Ratra\altaffilmark{1} and Michael S.\ Vogeley\altaffilmark{2}}

\altaffiltext{1}{Department of Physics, Kansas State University, 116 
                Cardwell Hall, Manhattan, KS 66506; email: ratra@phys.ksu.edu}
\altaffiltext{2}{Department of Physics, Drexel University, 3141 Chestnut 
                Street, Philadelphia, PA 19104; email: vogeley@drexel.edu}

\begin{abstract}
We review the current standard model for the evolution of the Universe
from an early inflationary epoch to the complex hierarchy of structure
seen today. We summarize and provide key references for the following
topics: observations of the expanding Universe; the hot early Universe
and nucleosynthesis; theory and observations of the cosmic microwave
background; Big Bang cosmology; inflation; dark matter and dark
energy; theory of structure formation; the cold dark matter model;
galaxy formation; cosmological simulations; observations of galaxies,
clusters, and quasars; statistical measures of large-scale structure;
and measurement of cosmological parameters. We conclude with
discussion of some open questions in cosmology. This review is
designed to provide a graduate student or other new worker in the
field an introduction to the cosmological literature.
\end{abstract}


\section{Introduction}

It is the current opinion of many physicists that the Universe is
well described by what Fred Hoyle termed a Big Bang Model, in which
the Universe expanded from a denser hotter childhood to its current
adolescence, with a present energy budget dominated by dark energy and
less so by dark matter, neither of which have been detected in the
laboratory, with the stuff biological systems, planets, stars, 
and all visible matter are made of (called baryonic matter by 
cosmologists) being a very small tracer on this dark sea, and with 
electromagnetic radiation being an even less significant contributor. 
Galaxies and groups and clusters of galaxies are locally distributed 
inhomogeneously in space, but on large enough scales and in a statistical 
sense the distribution approaches isotropy. This is supported by other
electromagnetic distributions such as the X-ray and cosmic microwave
backgrounds, which are close to isotropic.  As one looks out
further into space, as a consequence of the finite speed of light, one
sees objects as they were at earlier times, and there is clear 
observational evidence for temporal evolution in the distribution of various 
objects such as galaxies. 

At earlier times the Universe was hotter and denser, at some stage so 
hot that atoms could not exist. Nuclear physics reactions between 
protons, neutrons, etc., in the cooling expanding Universe resulted 
in the (nucleo)synthesis of the lighter elements (nuclei) such as D, 
$^4$He, and $^7$Li, with abundances in good accord with what is observed, 
and with the photons left over forming a residual cosmic microwave 
background (CMB) also in good agreement with what is observed. 

Given initial inhomogeneities in the mass distribution at an earlier time, 
processing of these by the expansion of the Universe, gravitational 
instability, pressure gradients, and microphysical processes, gives 
rise to observed anisotropies in the CMB
and the current large-scale distribution of nonrelativistic matter; the
situation on smaller spatial scales, where galaxies form, is
murkier. Observations indicate that the needed initial inhomogeneities
are most likely of the special form known as scale invariant, and that
the simplest best-fitting Big Bang Model has flat spatial geometry.  These
facts could be the consequence of a simple inflationary infancy of the 
Universe, a very early period of extremely rapid expansion, which stretched
zero-point quantum-mechanical fluctuations to larger length scales and
transmuted them into the needed classical inhomogeneities in the 
mass-energy distribution.  At the end of the inflationary expansion all 
radiation and matter is generated as the Universe moves into the usual 
Big Bang Model epoch. Inflation has roots in models of very high-energy physics.  Because of electromagnetic charge screening, 
gravity is the dominant large-scale force. General relativity is the 
best theory of gravity. 

This review attempts to elaborate on this picture. Given the Tantalus
principle of cosmology (and most of astrophysics), that one can see
but not ``touch" --- which makes this a unique field of physics ---
there have been many false starts and even much confusion and many
missed opportunities along what most now feel is the right
track. Given space constraints we cannot do justice to what are now
felt to be false starts, nor will we discuss more than one or two
examples of confusion and missed opportunities. We attempt here to
simply describe what is now thought to be a reasonable standard model
of cosmology and trace the development of what are now felt to be the
important threads in this tapestry; time will tell whether our use of
``reasonable standard" is more than just youthful arrogance (or
possibly middle-aged complacence).

In the following sections we review the current standard model of
cosmology, with emphasis in parts on some historical roots, citing
historically significant and more modern papers as well as review
articles. We begin with discussion of the foundations of the Big Bang
Model in Sec.\ \ref{foundations}, which summarizes research in the
half century from Einstein's foundational paper on modern cosmology
until the late 1960's discovery of the CMB radiation, as well as some
loose ends. Section \ref{inflation} discusses inflation, which
provides an explanation of the Big Bang that is widely felt to be
reasonable. Dark energy and dark matter, the two (as yet not directly
detected) main components of the energy budget of the present Universe
are reviewed in Sec.\ \ref{dark}. Further topics include the growth of
structure in the Universe (Sec.\ \ref{growth}), observations of
large-scale structure in the Universe (Sec.\ \ref{mapping}), and
estimates of cosmological parameters (Sec.\ \ref{parameters}). We
conclude in Sec.\ \ref{questions} with a discussion of what are now
thought to be relevant open questions and directions in which the
field appears to be moving.

We use hardly any mathematical equations in this review. In some cases
this results in disguising the true technical complexity of the issues
we discuss.

We exclude from this review a number of theoretical topics: quantum
cosmology, the multiverse scenario, string gas cosmology, braneworld
and higher dimensional scenarios, and other modifications of the
Einstein action for gravity. (We note that one motivation for
modifying Einstein's action is to attempt to do away with the
construct of dark matter and/or dark energy.  While it is probably too
early to tell whether this can get rid of dark energy, it seems
unlikely that this is a viable way of getting around the idea of dark
matter.)
 
For original papers written in languages other than English, we cite
only an English translation, unless this does not exist. We only 
cite books that are in English. For books that have been reprinted we 
cite only the most recent printing of which we are aware.

As a supplement to this review, we have compiled lists of key
additional reference materials and links to Web resources that will be
useful for those who want to learn more about this vast topic. These
materials, available on our Web site\footnote{See our Web site at {\tt
www.physics.drexel.edu/universe/}},
include lists of more technical books (including
standard textbooks on cosmology and related topics), historical and
biographical references, less technical books and journal articles,
and Web sites for major observatories and satellites.

\section{Foundations of the Big Bang Model}
\label{foundations}

\subsection{General relativity and the expansion of the Universe}
\label{GR}

Modern cosmology begins with \citet{Einstein:1917ab} where he applies
his general relativity theory to cosmology. At this point in time our Galaxy,
the Milky Way, was thought by most to be the Universe.  To make
progress Einstein assumed the Universe was spatially homogeneous and
isotropic; this was enshrined as the ``Copernican" cosmological
principle by \citet{Milne:1933ab}. \citet[][Sec.\ 3]{Peebles:1993ab}
reviews the strong observational evidence for large-scale statistical
isotropy; observational tests of homogeneity are not as
straightforward.  Einstein knew that the stars in the Milky Way moved
rather slowly and decided, as everyone had done before him, that the
Universe should not evolve in time. He could come up with a static
solution of his equations if he introduced a new form of energy, now
called the cosmological constant. It turns out that Einstein's static
model is unstable. In the same year \citet{deSitter:1917ab} found the
second cosmological solution of Einstein's general relativity
equations; \citet{Lemaitre:1925ab} and \citet{Robertson:1928ab}
re-expressed this solution in the currently more familiar form of the
exponentially expanding model used in the inflation
picture. \citet{Weyl:1923ab} noted the importance of prescribing
initial conditions such that the particle geodesics diverge from a
point in the past. \citet{Friedmann:1922ab, Friedmann:1924ab}, not
bound by the desire to have a static model, discovered the evolving
homogeneous solutions of Einstein's equations; \citet{Lemaitre:1927ab}
rediscovered these ``Friedmann-Lema\^\i tre"
models. \citet{Robertson:1929ab} initiated the study of metric tensors
of spatially homogeneous and isotropic spacetimes, and continuing
study by him and A.~G.~Walker (in the mid 1930's) led to the
``Robertson-Walker" form of the metric tensor for homogeneous world
models. Of course, in the evolving cosmological model solutions only
observers at rest with respect to the expansion/contraction see an
isotropic and homogeneous Universe; cosmology thus re-introduces
preferred observers!  \citet{North:1990ab} and \citet{Longair:2006ab}
provide comprehensive historical reviews.

See the standard cosmology textbooks for the modern formalism.

\subsection{Galaxy Redshift and Distance Measurements}
\label{redshift}

Meanwhile, with first success in 1912,
\citet{Slipher:1917ab}\footnote{ Although the ``canals" on Mars are
not really canals, they had an indirect but profound influence on
cosmology. Percival Lowell built Lowell Observatory to study the Solar
System and Mars in particular, and closely directed the research of
his staff. Slipher was instructed to study M31 and the other white
nebulae under the hope that they were proto-solar-systems.}  finds
that most of the ``white spiral nebulae" (so-called because they have
a continuum spectrum; what we now term spiral galaxies) emit light
that is redshifted (we now know that the few, including M31
(Andromeda) and some in the Virgo cluster, that emit blue-shifted
light are approaching us), and \citet{Eddington:1923ab} identifies
this with a redshift effect in the \citet{deSitter:1917ab} model (not
the cosmological redshift effect). \citet{Lemaitre:1925ab} and
\citet{Robertson:1928ab} derive Hubble's velocity-distance law $v =
H_0 r$ (relating the galaxy's speed of recession $v$ to its distance
$r$ from us, where $H_0$ is the Hubble constant, the present value of
the Hubble parameter) in the Friedmann-Lema\^\i tre models.  The
velocity-distance Hubble law is a consequence of the cosmological
principle, is exact, and implies that galaxies further away than the
current Hubble distance $r_H = c/H_0$ are moving away faster than the
speed of light $c$. \citet{Hubble:1925ab}\footnote{ Duncan had earlier
found evidence for variable stars in M33, the spiral galaxy in
Triangulum.}  uses Leavitt's \citep{Leavitt:1912ab,
Johnson:2005ab}\footnote{ Leavitt published a preliminary result in
1908 and Hertzsprung and Shapley helped develop the relation, but it
would be another 4 decades (1952) before a reasonably accurate version
became available (which led to a drastic revision of the distance
scale).}  quantitative Cepheid variable star period-luminosity
relation to establish that M31 and M33 are far away \citep[confirming
the earlier somewhat tentative conclusion of][]{Opik:1922ab}, and does
this for more galaxies, conclusively establishing that the white
nebulae are other galaxies outside our Milky Way galaxy (there was
some other earlier observational evidence for this position but
Hubble's work is what convinces people). Hubble gets Humason (middle
school dropout and one time muleskinner and janitor) to re-measure
some Slipher spectra and measure more spectra, and
\citet{Hubble:1929ab}\footnote{ In the mid 1920's Lundmark and
Str\"omberg had already noted that more distant galaxies seemed to
have spectra that were more redshifted.}  establishes Hubble's
redshift-distance law $cz = H_0 r$, where the redshift $z$ is the
fractional change in the wavelength of the spectral line under study
(although in the paper Hubble calls $cz$ velocity and does not mention
redshift). The redshift-distance Hubble law is an approximation to the
velocity-distance law, valid only on short distances and at low
redshifts.  \citet{North:1990ab} provides a comprehensive historical
review; \citet{Berendzen:1976ab} and \citet{Smith:1982ab} are more
accessible historical summaries. 
See the standard cosmology textbooks for the modern formalism. 
\citet{Branch:1998ab}
discusses the use of type Ia supernovae as standard candles for
measuring the Hubble constant.  See Fig.\ 1 of
\citet{Leibundgut:2001ab} for a recent plot of the Hubble
law. \citet{Harrison:1993ab}, \citet{Davis:2004ab}, and
\citet{Lineweaver:2005ab} provide pedagogical discussions of
issues related to
galaxies moving away faster than the speed of light.

\subsection{The Hot Early Universe and Nucleosynthesis}
\label{nucleosynthesis}

As one looks out further in space (and so back in time, because light
travels at finite speed) wavelengths of electromagnetic radiation we 
receive now have been redshifted further by the expansion and so
Wien's law tells us (from the blackbody CMB) that the temperature was
higher in the past. The younger Universe was a hotter, denser place.
Lema\^\i tre (``the father of the Big Bang") emphasized the importance of
accounting for the rest of known physics in the general relativistic
cosmological models. 

Early work on explaining the astrophysically
observed abundances of elements assumed that they were a consequence
of rapid thermal equilibrium reactions and that a rapidly falling
temperature froze the equilibrium abundances. Tolman, Suzuki, von
Weizs\"acker, and others in the 1920's and 1930's argued that the observed 
helium-hydrogen ratio in this scenario required that at some point the 
temperature had been at least $10^9$ K (and possibly as much as $10^{11}$ K).
\citet{Chandrasekhar:1942ab} performed the first
detailed, correct equilibrium computation and concluded that no single
set of temperature and density values can accommodate all the observed
abundances; they suggested that it would be useful to consider a
non-equilibrium process. \citet{Gamow:1946ab}, building 
on his earlier work, makes the
crucial point that in the Big Bang Model ``the conditions necessary for 
rapid nuclear reactions were existing only for a very short time, so that 
it may be quite dangerous to speak about an equilibrium state", i.e., the 
Big Bang was the place to look for this non-equilibrium process. 

\citet{Gamow:1948ab}, a student of Friedmann, and 
\citet{Alpher:1948ab}, a student of Gamow, estimated the radiation 
(photon) temperature at nucleosynthesis, and from the Stefan-Boltzmann 
law for blackbody radiation noted that the energy budget of the Universe 
must then have been dominated by radiation. \citet{Gamow:1948ab} 
evolved the radiation to the much later epoch of matter-radiation equality 
(the matter and radiation energy densities evolve in different ways and 
this is the time at which both had the same magnitude), a concept also 
introduced by Gamow, while \citet{Alpher:1948cd} predicted 
a residual CMB radiation at the present time from nucleosynthesis and 
estimated its present temperature to be 5 K (because the zero-redshift
baryon density was not reliably known then, it is somewhat of a
coincidence that this temperature estimate is close to the observed 
modern value). \citet{Hayashi:1950ab}
pointed out that at temperatures about 10 times higher than during
nucleosynthesis rapid weak interactions lead to a thermal equilibrium
abundance ratio of neutrons and protons determined by the neutron-proton 
mass difference, which becomes frozen in as the expansion decreases the 
temperature, thus establishing the initial conditions for nucleosynthesis.
This is fortunate, in that an understanding of higher energy physics is
not needed to make firm nucleosynthesis predictions; this is also
unfortunate, because element abundance observations cannot be used to 
probe higher energy physics. 

\citet{Alpher:1953ab}
conclude the early period of the standard model of nucleosynthesis. By
this point it was clear that initial hopes to explain all observed
abundances in this manner must fail, because of the lack of stable
nuclei at mass numbers 5 and 8 and because as the temperature drops 
with the expansion it becomes more difficult to penetrate the 
Coulomb barriers. Cosmological nucleosynthesis can only generate the 
light elements and the heavier elements are generated from these 
light elements by further processing in the stars.

\citet{Zeldovich:1963ab} and \citet{Smirnov:1964ab} 
noted that the $^4$He and D abundances are sensitive to the baryon density:
the observed abundances can be used to constrain the baryon density. 
\citet{Hoyle:1964ab} carried out a detailed computation 
of the $^4$He abundance and on comparing to measurements concluded ``most, 
if not all, of the material of our .... Universe, has been `cooked' to a
temperature in excess of $10^{10}$ K". They were the first to note that
the observed light element abundances were sensitive to the expansion
rate during nucleosynthesis and that this could constrain new physics at
that epoch (especially the number of light, relativistic, neutrino
families). 

After Penzias and Wilson measured the CMB (see below), 
\citet{Peebles:1966ab, Peebles:1966cd} computed the abundances of 
D, $^3$He, and $^4$He, and their dependence on, among other things, the 
baryon density and the expansion rate during nucleosynthesis. The 
monumental \citet{Wagoner:1967ab} paper established the 
ground rules for future work. For a history of these developments see 
pp.~125-128 and 240-241 of \citet{Peebles:1971ab}, the articles
by Alpher and Herman and Wagoner on pp.\ 129-157 and 159-185, 
respectively, of \citet{Bertotti:1990ab}, and Ch.\ 3 and Sec.\ 7.2 
of \citet{Kragh:1996ab}. 
See the standard cosmology textbooks for the modern formalism. 
Accurate abundance predictions require 
involved numerical analysis; on the other hand pedagogy could benefit 
from approximate semi-analytical models
\citep{Bernstein:1989ab, Esmailzadeh:1991ab}. 

In the simplest nucleosynthesis scenario, the baryon density estimated
from the observed D abundance is consistent with that estimated from
WMAP CMB anisotropy data, and higher than that estimated from the
$^4$He and $^7$Li abundances. This is further discussed in
Sec. \ref{constraints}. \citet{Field:2006ab}, \citet{Cyburt:2004ab},
and \citet{Steigman:2006ab} are recent reviews of nucleosynthesis.

In addition to residual CMB radiation, there is also a residual neutrino 
background. Above a temperature of about $10^{10}$ K the CMB photons have
enough energy to produce a thermal equilibrium abundance of neutrinos.
Below this temperature the neutrinos decouple and freely expand, resulting
in about 300 neutrinos per cubic centimeter now (with three families, and 
this number also includes antineutrinos), at a temperature of about 2 K, 
lower than that of the CMB because electron-positron annihilation heats the 
CMB a little. See \citet{Dolgov:2002ab}, \citet{Hannestad:2006ab}, and the 
more recent textbooks cited below for more detailed discussions of the 
(as yet undetected) neutrino background. We touch on neutrinos again in 
Sec. \ref{gravity}.

\subsection{Theory and Observations of the CMB}
\label{CMB}

The CMB radiation contributes of the order of 1\% of the static or ``snow"
seen when switching between channels on a television with a
conventional VHF antenna; it is therefore not surprising that it had been
detected a number of times before its 1965 discovery/identification.  
For instance, it is now known that \citet{McKellar:1941ab} 
deduced a CMB temperature of 2.3 K at a wavelength of 2.6 mm by estimating 
the ratio of populations in the first excited rotational and ground states
of the interstellar cyanogen (CN) molecule (determined from absorption 
line measurements of Adams). It
is now also known that the discrepancy of 3.3 K between the measured
and expected temperature of the Bell Labs horn antenna (for
communicating with the Echo I satellite) at a wavelength of 12.5 cm
found by \citet{Ohm:1961ab} is due to the CMB. Ohm also
notes that an earlier measurement with this
telescope \citep{DeGrasse:1959ab} ascribes a temperature of $2 \pm 1$ K
to back and side lobe pick up, that this is ``...temperature not
otherwise accounted for...", and that ``it is somewhat larger than
the calculated temperature expected".  Of course, McKellar had the
misfortune of performing his analyses well before Gamow and
collaborators had laid the nucleosynthesis foundations that would
eventually explain the CN measurements (and allow the CMB interpretation) 
and Ohm properly did not overly
stress the discrepancy beyond its weak statistical significance. 

While Alpher and Herman [e.g., pp.~114-115 of \citet{Alpher:2001ab}
and p.~130 of \citet{Weinberg:1993ab}] privately raised the issue of
searching for the CMB, and Hoyle came close to correctly explaining
McKellar's CN measurements [see pp.~345-346 of \citet{Kragh:1996ab}],
\citet[p.\ 491, p.\ 89, and p.\ 315, respectively]{Zeldovich:1963ab,
Zeldovich:1963cd, Zeldovich:1965ab}, \citet{Doroshkevich:1964ab} and
\citet[p.\ 448]{Dicke:1965ab} are the first published discussions of
possible observational consequences of the (then still hypothetical)
CMB in the present Universe. The relevant discussions of Zel'dovich
and Doroshkevich \& Novikov are motivated by the same nucleosynthesis
considerations that motivated Gamow and collaborators; Dicke and
Peebles favored an oscillating Universe and needed a way to destroy
heavy elements from the previous cycle and so postulated an initial
hotter stage in each cycle. Both \citet{Doroshkevich:1964ab} and
\citet{Zeldovich:1965ab} refer to \citet{Ohm:1961ab} but neither
appear to notice Ohm's 3.3 K discrepancy; in fact \citet{Zeldovich:1965ab}
(incorrectly) argues that Ohm constrains the temperature to be less
than 1 K and given the observed helium abundance this rules out the
hot Big Bang Model!

Working with the same antenna as Ohm, using the Dicke switching technique
to compare the antenna temperature to a liquid helium load at a known
temperature, and paying very careful attention to possible systematic 
effects, \citet{Penzias:1965ab} measure the excess temperature to be 
$3.5 \pm 1$ K at 7.35 cm wavelength; \citet{Dicke:1965cd} identify this 
as the CMB radiation left over from the hot Big Bang. 

The CMB is the dominant component of the radiation density of the 
Universe, with a density now of about 400 CMB photons per cubic centimeter 
at a temperature of about 2.7 K now.  As noted in the previous subsection,
observed light element abundances in conjunction with nucleosynthesis
theory allows for constraints on the density of baryonic matter.
Thus, there are a few billion CMB photons for every baryon; the CMB
photons carry most of the cosmological entropy.

To date there is no observational indication of any deviation of the
CMB spectrum from a Planckian blackbody.  \citet{Partridge:1995ab} 
reviews early measurements of the CMB spectrum.  A definitive 
observation of the CMB spectrum was made by COBE (see \citet{Gush:1990ab}
for a contemporaneous rocket-based measurement), which measured a 
temperature of $2.725\pm0.002$ K (95~\% confidence) \citep{Mather:1999ab}
and 95~\% confidence upper limits on possible spectral distortions: 
$|\mu|<9\times 10^{-5}$ for the chemical potential of early ($10^5 
< z < 3\times 10^6$) energy release and $|y|<1.5\times 10^{-5}$ for 
Comptonization of the spectrum at later times \citep{Fixsen:1996ab}. 
\citet{Wright:1994ab} shows that these constraints strongly 
rule out many alternatives to the Big Bang Model, including the steady 
state model and explosive galaxy formation.

Anisotropy of the CMB temperature, first detected by COBE
\citep{Smoot:1992ab}, reveals important features of the formation and
evolution of structure in the Universe.  A small dipole anisotropy
(discovered in the late 1960's and early 1970's by Conklin and Henry
and confirmed by Corey and Wilkinson as well as Smoot, Gorenstein, and
Muller) is caused by our peculiar motion; the CMB establishes a
preferred reference frame.  Higher multipole anisotropies in the CMB
reflect the effect of primordial inhomogeneities on structure at the
epoch of recombination and more complex astrophysical effects along
the past light cone that alter this primordial anisotropy. We discuss
these anisotropies, as well as the recently-detected polarization
anisotropy of the CMB in Sec.\ \ref{anisotropy}. The anisotropy signal
from the recombination epoch allows precise estimation of cosmological
parameters (see Sec.\ \ref{constraints}).

In addition to references cited above, \citet[pp.\ 64-70]{Dicke:1970ab}, 
\citet{Wilson:1983ab}, \citet{Wilkinson:1990ab}, 
\citet[Ch.\ 2]{Partridge:1995ab}, and \citet[Sec.\ 7.2]{Kragh:1996ab} review 
the history.  For the modern formalism see the more recent standard cosmology 
textbooks and \citet{Kamionkowski:1999ab}.

\subsection{Challenges for the Big Bang Model}
\label{challenges}

Since the Universe is now expanding, at earlier times it was denser
and hotter. A naive extrapolation leads to a (mathematical) singularity 
at the beginning, with infinite density and temperature, at the initial
instant of time, and over all space. This naive extrapolation is 
unjustified since the model used to derive it breaks down physically
before the mathematical singularity is reached. Deriving the
correct equations of motion for the very early Universe is an
important area of current research. While there has been much work,
there is as yet no predictive model that unifies gravity and quantum
mechanics --- and this appears essential for an understanding of the
very early Universe, because as one goes back in time the
gravitational expansion of the Universe implies that large cosmological
length scales now correspond to tiny quantum mechanical scales in the
very very early Universe. There is a small but active group of workers
who believe that only a resolution of this issue (i.e., the derivation
of a full quantum theory of gravity) will allow for progress on the
modeling of the very early Universe. But most others, perhaps
inspired by the wonderful successes of particle physics models that
have successfully described shorter and shorter distance
physics, now believe that it is important to try to solve some of the
``problems" of the Big Bang Model by attempting to model the
cosmophysical world at an energy density higher than is probed by
nucleosynthesis and other lower redshift physics, but still well below
the Planck energy density where quantum gravitational effects are
important. This is the approach we take in the following
discussion, by focusing on ``problems" that could be resolved below the
Planck density. Whether Nature has chosen this path is as yet unclear,
but at least the simplest versions of the inflation scenario
(discussed in the next section) are compatible with current observations 
and will likely be well tested by data acquired within this decade.

Assuming just nonrelativistic matter and radiation (CMB and neutrinos) in
order of magnitude agreement with observations, the distance over
which causal contact is possible grows with the age of the Universe.
That is, if one assumes that in this model the cosmological principle is
now valid because of ``initial conditions" at an earlier time, then
those initial conditions must be imposed over distances larger than
the distance over which causal communication was possible. (And maybe this is
what a quantum theory of gravitation will do for cosmology, but in the
spirit of the earlier discussion we will view this as a ``problem" of
the Big Bang Model that should be resolved by physics at energies
below the Planck scale.)  \citet[p.\ 1349]{Alpher:1953ab}
contains the earliest remarks (in passing) that we are aware
of about this particle horizon problem. The terminology is due to
\citet{Rindler:1956ab} which is an early discussion of 
horizons in general. \citet{Harrison:1968ab} also mentions the 
particle horizon problem in passing, but \citet{McCrea:1968ab}
and \citet{Misner:1969ab} contain the first clear statements 
we are aware of, with Misner stating ``These Robertson-Walker models 
therefore give no insight into why the observed microwave radiation from 
widely different angles in the sky has ... very precisely ... the same 
temperature".  Other early discussions are in \citet[p.\ 61]{Dicke:1970ab}, 
\citet{Doroshkevich:1970ab}, and the text books of 
\citet[pp.\ 525-526]{Weinberg:1972ab} and \citet[pp.\ 815-816]{Misner:1973ab}, 
This issue was discussed in many papers and books starting in the early 
1970's, but the celebrated \citet{Dicke:1979ab} review is often credited 
with drawing prime-time attention to the particle horizon ``problem".

The large entropy of the Universe (as discussed above, there are now a
few billion CMB photons for every baryon) poses another puzzle. When
the Universe was younger and hotter there had to have been a thermal
distribution of particles and antiparticles and, as the Universe
expanded and cooled, particles and antiparticles annihilated into
photons, resulting in the current abundance of CMB photons and
baryons. Given the lack of a significant amount of antibaryons now,
and the large photon to baryon ratio now, at early times there must
have been a slight (a part in a few billion) excess of baryons over
antibaryons. We return to this issue in the next section.

\section{Inflation}
\label{inflation}

It is possible to trace a thread in the particle horizon problem
tapestry back to the singularity issue and early discussions of
Einstein, Lema{\^\i}tre, and others who viewed the singularity as arising
from the unjustified assumption of exact isotropy, and led to the
intensive study of homogeneous but anisotropic cosmological models in
the late 1960's and early 1970's. These attempts failed to tame the singularity
but did draw attention to isotropy and the particle horizon problem of
the standard Big Bang Model.  It is interesting that this singularity
issue also drove the development of the steady state picture, which in
its earliest version was just a de Sitter model. While observations
soon killed off the original steady state model \citep[a more recent variant, 
the quasi-steady state model, can be adjusted to accommodate the data, 
see, e.g.,][]{Narlikar:2003ab}, the idea of a possible early,
pre-Big-Bang, nonsingular de Sitter epoch thrived. It appears that 
\citet{Brout:1978ab} were the first to note that such a 
cosmological model was free of a particle horizon. However, they do not 
seem to make the connection that this could allow for isotropy by ensuring 
that points well separated now shared some common events in the past and 
thus causal physics could in principle make the Universe isotropic.  
\citet{Zee:1980ab} noted that if one modifies the early Universe by 
speeding up the expansion rate enough over the expansion rate during the 
radiation dominated epoch, the particle horizon problem is resolved (but 
he does not go to the exponentially expanding de Sitter solution 
characteristic of the early inflation scenario). 

\citet{Sato:1981ab,
Sato:1981cd}, \citet{Kazanas:1980ab}, and \citet{Guth:1981ab} 
are the ones who make the (now viewed to be crucial) point that during a 
phase transition at very high temperature in grand unified models it is 
possible for the grand unified Higgs scalar field energy density to 
behave like a cosmological constant, driving a de Sitter exponential 
cosmological expansion, which results in a 
particle-horizon-free cosmological model. And the tremendous expansion 
during the de Sitter epoch will smooth out wrinkles in the matter 
distribution, by stretching them to very large scales, an effect alluded 
to earlier by \citet{Hoyle:1962ab} in the context of the 
steady state model, which could result in an isotropic Universe now, 
provided the initial wrinkles satisfy certain conditions. See 
\citet{Ellis:1988ab} and \citet{Narlikar:1991ab} for caveats and criticism. Of
course, to get the inflationary expansion started requires a large
enough, smooth enough initial patch. The contemporary explanation
appeals to probability: loosely, such a patch will exist somewhere and
inflation will start there. In addition the initial conditions issue
is not completely resolved by inflation, only greatly alleviated;
since inflation stretches initially small length scales to length
scales of contemporary cosmological interest, the cosmological
principle requires that there not be very large irregularities on very
small length scales in the very early Universe. This could be a clue
to what might be needed from a model of very high energy,
pre-inflation, physics. For reviews of inflation see the more recent 
standard cosmology and astroparticle physics textbooks.

Building on ideas of Brout, Englert, and collaborators, 
\citet{Gott:1982ab} noted that it was possible to have inflation result 
in a cosmological model with open spatial hypersurfaces at the present 
time, in contrast to the Sato-Kazanas-Guth discussion that focused on 
flat spatial hypersurfaces. This open-bubble inflation model, in which 
the observable part of the contemporary Universe resides inside a bubble
nucleated (because of a small upward ``bump" in the potential energy
density function) between two distinct epochs of inflation, is a clear
counter-example to the oft-repeated (but incorrect) claim that 
inflation explains why the Universe appears to have negligible space 
curvature. See \citet{Ratra:1994ab, Ratra:1995ab} for a more 
detailed discussion of this model. 

The open-bubble inflation model was the first consistent inflation
model.  Unfortunately for the Guth model, as the phase transition
completes and one hopes to have a smooth transition to the more
familiar radiation-dominated expansion of the hot Big Bang Model, one
finds that the potential in the Guth model results in many small
bubbles forming with most of energy density residing in the bubble
walls. In this model the Universe at the end of inflation was very
inhomogeneous because the bubble collisions were not rapid enough to
thermalize the bubble wall energy density (i.e., the bubbles did not
``percolate").  \citet{Linde:1982ab} and \citet{Albrecht:1982ab} used
a specific potential energy density function for the Higgs field in a
grand unified model and implemented Gott's scenario in the
Sato-Kazanas-Guth picture, except they argued that the second epoch of
inflation lasts much longer than Gott envisaged and so stretches the
bubble to length scales much larger than the currently observable part
of the Universe, thus resulting in flat spatial hypersurfaces now. The
great advantage of the Gott scenario is that it uses the first epoch
of inflation to resolve the particle horizon/homogeneity problem and
so this problem does not constrain the amount of inflation after the
bubble nucleates. \citet{Brout:1978ab} and \citet{Coleman:1980ab} note
that symmetry forces the nucleating bubble to have an open geometry,
and this is why inflation requires open spatial hypersurfaces, but
with significant inflation after bubble nucleation the radius of
curvature of these hypersurfaces can be huge. Thus the amount of space
curvature in the contemporary Universe is a function of the amount of
inflation after bubble nucleation, and it is now widely accepted that
observational data (as discussed below in Sec.\ \ref{constraints}) are
consistent with an insignificant amount of space curvature and thus
significant inflation after bubble nucleation.

It is well known that phase transitions can create topological
defects.  Grand unified phase transitions are no exception and often
create monopoles and other topological defects. If the Universe is
also inflating through this phase transition then the density of such
topological defects can be reduced to levels consistent with the
observations. This is not another argument in support of inflation,
although it is often claimed to be: it is just a way of using
inflation to make viable a grand unified theory that is otherwise
observationally inconsistent.

One major motivation for grand unification is that it allows for a
possible explanation of the observed excess of matter over antimatter
(or the baryon excess) mentioned in the previous section. There are
other possible explanations of how this baryon excess might have come
about. One much discussed alternative is the possibility of forming it
at the much lower temperature electroweak phase transition, through a
non-perturbative process, but this might raise particle
horizon or homogeneity issues. However, at present there is no
convincing, numerically satisfying explanation of the baryon excess,
from any process.  \citet{Quinn:1998ab}, \citet{Dine:2004ab},
\citet{Trodden:2004ab}, and \citet{Cline:2006ab} review models now
under discussion for generating the baryon excess.

At the end of inflation, as the phase transition completes and the 
Universe is said to reheat, one expects the generation of matter and 
radiation as the Universe makes the transition from rapid inflationary 
expansion to the more sedate radiation-dominated expansion of the hot 
Big Bang Model. This is an area of ongoing research and it would be 
useful to have a convincing, numerically satisfying 
model of this epoch. The baryon excess might be generated during this
reheating process. 

While great effort has been devoted to inflation, resulting in a huge 
number of different models, at the present stage of development inflation 
is a very interesting general scenario desperately in need 
of a more precise and more convincing very high energy particle physics 
based realization. As far as large-scale cosmology is concerned, inflation
in its simplest form is modeled by a scalar field (the inflaton) whose
potential energy density satisfies certain properties that result in a 
rapid enough cosmological expansion at early times. It is interesting 
that cosmological observations within this decade might firm up this 
model of the very early Universe based on very high energy physics before 
particle physicists do so. For reviews see the more recent standard 
cosmology and astroparticle physics textbooks.

Assuming an early epoch of inflation, the cumulative effect of the 
expansion of the Universe from then to the present means that contemporary 
cosmological length scales (e.g., the length scale that characterizes the 
present galaxy distribution) correspond to very tiny length scales 
during inflation, so tiny that quantum-mechanical zero-point fluctuations 
must be considered in any discussion involving physics on these length scales.

As mentioned above, the idea of an early de-Sitter-like expansion
epoch, pre-Big-Bang, was discussed in the 1970's, as a possible way of
taming the initial singularity. While this de Sitter epoch was typically
placed at very high energy, it differs significantly from the
inflation scenario in that it was not driven by a scalar field
potential energy density. Nevertheless because it was at energies close to the
Planck energy there were many discussions of quantum mechanical
fluctuations in de Sitter spacetime in the 1970's.

In the inflation case quantum mechanics introduces additional fluctuations,
the zero-point fluctuations in the scalar field. This was noted by 
\citet{Hawking:1982ab}, \citet{Starobinsky:1982ab}, and \citet{Guth:1982ab}, 
and further studied by \citet{Bardeen:1983ab}. For a discussion of scalar 
field quantum fluctuations in de Sitter spacetime and their consequences 
see \citet{Ratra:1985ab}. \citet{Fischler:1985ab} use the 
Dirac-Wheeler-DeWitt formalism to consistently semi-classically quantize 
both gravitation and the scalar field about a de Sitter background, and 
carry through a computation of the power spectrum of zero-point fluctuations. 
The simplest inflation models have a weakly coupled scalar field and so a 
linear perturbation theory computation suffices. The fluctuations obey 
Gaussian statistics and so can be completely characterized by their two-point 
correlation function or equivalently their power spectrum. Inflation models 
that give non-Gaussian fluctuations are possible (for a review see 
\citet{Bartolo:2004ab}), but the observations do not yet demand this, 
being almost completely consistent with Gaussianity (see discussion in 
Sec.\ \ref{anisotropy} below). The simplest models give adiabatic or 
curvature (scalar) fluctuations; these are what result from adiabatically 
compressing or decompressing parts of an exactly spatially homogeneous 
Universe. More complicated models of inflation can produce fluctuations 
that break adiabaticity, such as (tensor) gravitational waves 
\citep{Rubakov:1982ab} and (vector) magnetic fields \citep{Turner:1988ab, 
Ratra:1992ab}, which might have interesting observational consequences 
(see Secs.\ \ref{gravity} and \ref{anisotropy} below). 

The power spectrum of energy density fluctuations depends on the model 
for inflation. If the scalar field potential energy density during 
inflation is close to flat and dominates the scalar field energy density, 
the scale factor grows exponentially with time (this is the de Sitter 
model), and after inflation but at high redshift the power spectrum of 
(scalar) mass-energy density fluctuations with wavenumber magnitude $k$ is 
proportional to $k$, or scale invariant, on all interesting length scales, 
i.e., curvature fluctuations diverge only as $\log k$. This was noted in 
the early 1980's for the inflation model \citep{Hawking:1982ab, 
Starobinsky:1982ab, Guth:1982ab}, although the virtues of a scale-invariant 
spectrum were emphasized in the early 1970's, well before inflation, by 
\citet{Harrison:1970ab}, \citet{Peebles:1970ab}, and \citet{Zeldovich:1972ab}. 
When the scalar field potential energy density is such that the scalar 
field kinetic energy
density is also significant during inflation a more general spectrum 
proportional to $k^n$ can result (where the spectral index $n$ depends 
on the slope of the potential energy density during inflation); for 
$n \not= 1$ the spectrum is said to be tilted \citep{Abbott:1984ab, 
Lucchin:1985ab, Ratra:1992cd}. Current observations appear to be 
reasonably well fit by $n = 1$. More complicated, non-power-law 
spectra are also possible.

We continue this discussion of fluctuations in Sec.\ \ref{growth} below.

\section{Dark Matter and Dark Energy}
\label{dark}

Most cosmologists are of the firm opinion that observations indicate
the energy budget of the contemporary Universe is dominated
by dark energy, with the next most significant contributor being 
dark matter, and with ordinary baryonic matter in a distant third place. 
Dark energy and dark matter are hypothetical constructs generated to 
explain observational data, and the current model provides a good, 
but not perfect, explanation of contemporary cosmological observations. 
However, dark energy and dark matter have not been directly detected 
(in the lab or elsewhere).

\citet{Hubble:1926ab}\footnote{
In this paper, among other things, Hubble also developed his galaxy 
classification scheme (of ellipticals, normal and barred spirals, 
and irregulars) and showed that the averaged large-scale galaxy 
distribution is spatially isotropic. \citet{Opik:1922ab} had 
earlier estimated the mass of M31.}
presented the first systematic 
estimate of masses of the luminous part of galaxies (based on studying 
the motion of stars in galaxies), as well as an estimate of the mass 
density of the Universe (using counts of galaxies in conjunction with 
the estimated masses of galaxies). 

Under similar assumptions (the validity of Newton's second law of 
motion and Newton's inverse-square law of gravitation, and that  
the large-scale structure under investigation is in gravitational 
equilibrium), \citet{Zwicky:1933ab}, in perhaps
one of the most significant discoveries of 
the previous century, found that galaxies in the Coma cluster of 
galaxies were moving with surprisingly high speeds. In 
modern terms, this indicates a Coma cluster mass density at least an order of 
magnitude greater than what would be expected from spreading the mass 
associated with the luminous parts of the galaxies in the Coma cluster
over the volume of the cluster. 
Zwicky's measurements probe larger 
length scales than Hubble's and so might be detecting mass that lies 
outside the luminous parts of the galaxies, i.e., mass that does not 
shine, or dark matter. Ordinary baryonic matter is largely 
nonrelativistic in the contemporary Universe and hence would be pulled 
in by the gravitational field of the cluster. Nucleosynthesis and CMB 
anisotropy measurements constrain the mass density of ordinary baryonic 
matter, and modern data indicate that not only is the amount of 
gravitating mass density detected in Zwicky-like observations 
significantly greater than what is shining, it is likely a factor of 3 to
5 times the mass density of ordinary baryonic matter. (It is also 
known that a large fraction of the expected baryonic matter can 
not significantly shine.) \citet{Smith:1936ab} confirmed 
Zwicky's result, using Virgo cluster measurements, and 
\citet{Zwicky:1937ab}\footnote{
In this paper Zwicky also proposes the remarkable idea of using 
gravitational lensing of background objects by foreground clusters 
of galaxies to estimate cluster masses.}
soon followed up with a more detailed paper.

Babcock's Ph.D.\ thesis \citep{Babcock:1939ab} was the next major 
(in hindsight) development in the dark matter story. He measured the 
rotation speed of luminous objects in or near the disk of the Andromeda 
(M31) galaxy, out to a distance of almost 20 kpc from 
the center and found that the rotation speed was still rising, not 
exhibiting the $1/\sqrt{r}$ Keplerian fall off with distance $r$ from
the center that would be expected if the mass distribution in M31 followed the 
distribution of the light. That is, Babcock found that the outer part of 
the luminous part of M31 was dominated by matter that did not shine. 
Soon thereafter \citet{Oort:1940ab} noted a similar result
for the galaxy NGC 3115. Almost two decades later,
\citet{vandeHulst:1957ab} confirmed Babcock's result by using 21 cm 
wavelength observations of hydrogen gas clouds that extend beyond the 
luminous part of M31, finding a roughly flat rotation curve at the edge 
(no longer rising with distance as Babcock had found). While there was 
some early theoretical discussion of this issue, the much more detailed 
M31 flat rotation curve measured by \citet{Rubin:1970ab} (Rubin was a 
student of Gamow) forced this dark matter into the limelight.

Other early indications of dark matter came from measurements of the
velocities of binary galaxies \citep{Page:1952ab} and the dynamics of
our Local Group of galaxies \citep{Kahn:1959ab}. 
\citet{deVaucouleurs:1969ab} and \citet{Arp:1969ab} found that the elliptical
galaxy M87 in the Virgo cluster had a faint mass-containing halo. 
\citet{Ostriker:1973ab} noted that one way of
making the disk of a spiral galaxy stable against a bar-like
instability is to embed it in a massive halo, and soon thereafter
\citet{Einasto:1974ab} and \citet{Ostriker:1974ab} showed that this 
suggestion was consistent with the observational evidence. These early 
results have been confirmed by a number of different techniques, 
including measuring the X-ray temperature of hot gas in galaxy clusters 
(which is a probe of the gravitational potential --- and the mass which 
generates it --- felt by the gas), and measurements of gravitational 
lensing of background sources by galaxy clusters. See Sec.\ \ref{constraints} 
for further discussion of this.

For reviews of dark matter see Sec.\ IV of \citet{Peebles:1971ab}
(note the fascinating comment on p.\ 64 on the issue of dark matter in 
clusters: ``This quantity" $M/L$ or the mass to luminosity ratio ``is 
suspect because when it is used to estimate the masses of groups or 
clusters of galaxies the result often appears to be unreasonable", i.e., 
large), \citet{Faber:1979ab}, \citet{Trimble:1987ab}, \citet{Ashman:1992ab},
\citet[][Sec.\ 18]{Peebles:1993ab}, and \citet{Einasto:2005ab}.

Much as van Maanen's measurements of large (but erroneous) rotation 
velocities for a number of galaxies prompted \citet{Jeans:1923ab}
to consider a modification of Newton's inverse-square law for gravity 
such that the gravitational force fell off slower with distance on large 
distances, the large (but not erroneous) velocities measured by Zwicky and 
others prompted \citet{Finzi:1963ab}, and many since then, 
to consider modifications of the law of gravity. The current observational 
indications are that this is not a very viable alternative to the dark 
matter hypothesis \citep[Secs.\ IV.A.1 and IV.B.13]{Peebles:2003ab}. 
In some cases, modern high energy physics suggests possible motivations 
for modifications of the inverse square law on various 
length scales; this is beyond the scope of our review.  

\citet{Milgrom:1983ab, Milgrom:2002ab} proposes a related but
alternate hypothesis: Newton's second law of motion is modified at low
accelerations. This hypothesis -- dubbed modified Newtonian dynamics
(MOND) -- does a remarkable job of fitting the flat rotation curves of
spiral galaxies, but most who have cared to venture an informed
opinion believe that it cannot do away completely with dark matter,
especially in low-surface-brightness dwarf galaxies and rich clusters
of galaxies.  More importantly, the lack of a well motivated extension
of the small-length-scale phenomenological MOND hypothesis that is
applicable on large cosmological length scales greatly hinders testing
the hypothesis. For a recent attempt at such an extension see
\citet{Bekenstein:2004ab}. For a preliminary sketch of cosmology in
this context see \citet{Diaz-Rivera:2006ab}. For a review of MOND see
\citet{Sanders:2002ab}.

Most cosmologists are convinced that dark matter
exists. Nucleosynthesis constraints indicate that most of the dark
matter is not baryonic.  \citep[Not all baryons shine; for a review of
options for dark baryons see][]{Carr:1994ab}. Galaxies are in general
older than larger-scale structures (such as clusters); this indicates
that the dark matter primeval velocity dispersion is small (for if it
were large gravity would be able to overcome the corresponding
pressure only on large mass --- and so length --- scales, first
forming large-scale objects that fragment later into younger
smaller-scale galaxies). Dark matter with low primeval velocity
dispersion is known as cold dark matter (CDM).  More precisely, the
CDM model assumes that most of the nonrelativistic matter-energy of
the contemporary Universe is in the form of a gas of massive,
non-baryonic, weakly-interacting particles with low primeval velocity
dispersion. One reason they must be weakly interacting is so they do
not shine. \citet{Munoz:2004ab}, \citet{Bertone:2005ab}, and
\citet{Baltz:2004ab} review particle physics dark matter candidates
and prospects for experimental detection. \citet{Bond:1982ab} and
\citet{Blumenthal:1982ab} note the advantages of CDM and that modern
high energy physics models provide plausible hypothetical candidates
for these particles.  \citet{Peebles:1982ab} casts the cosmological
skeleton of the CDM model, emphasizing that in this model structure
forms from the gravitational growth of primordial departures from
homogeneity that are Gaussian, adiabatic, and scale invariant,
consistent with what is expected from the simplest inflation
models. \citet{Blumenthal:1984ab} is a first fleshing out of the CDM
model. See \citet{Peebles:1993ab} and \citet{Liddle:2000ab} for
textbook discussions of the CDM model. More details about this model,
including possible problems, are given in Sec.\ \ref{growth} below.

To set the numerical scale for cosmological mass densities, following 
\citet{Einstein:1932ab}, one notes that the 
simplest Friedmann-Lema\^{\i}tre model relevant to the contemporary 
Universe is one with vanishing space curvature and with energy budget 
dominated by non-relativistic matter (and no cosmological constant). In 
this critical or Einstein-de Sitter case the Friedmann equation fixes 
the energy density of nonrelativistic matter for a given value of the 
Hubble constant. Cosmologists then define the mass-energy density 
parameter $\Omega$ for each type of mass-energy (including that of the 
curvature of spatial hypersurfaces $\Omega_{\rm K}$, the cosmological
constant $\Omega_\Lambda$, and nonrelativistic matter $\Omega_{\rm M}$) 
as the ratio of that mass-energy density to the critical or Einstein-de 
Sitter model mass-energy density. The Friedmann equation implies that 
the mass-energy density parameters sum to unity. (In general the 
$\Omega$'s are time dependent; in what follows numerical values 
for these parameters refer to the current epoch.)

As discussed in Sec.\ \ref{constraints} below, it has long been known
that nonrelativistic matter (baryons and CDM) contributes about 25 or
30 \% to the critical mass-energy density. After the development of
the inflation picture for the very early Universe in the 1980's there
was a wide-spread belief that space curvature could not contribute to
the mass-energy budget (this is not necessary, as discussed above),
and for this and a few other reasons (among others, the time scale
problem arising from the large measured values of the Hubble constant
and age of the Universe), \citet{Peebles:1984ab} proposed that
Einstein's cosmological constant contributed the remaining 70 or 75 \%
of the mass-energy of the Universe. This picture was soon generalized
to allow the possibility of a scalar- field energy density that is
slowly varying in time and close to homogeneous in space, what is now
called dark energy \citep{Peebles:1988ab, Ratra:1988ab}. As discussed
in Sec.\ \ref{constraints} below, these models predict that the
expansion of the Universe is now accelerating and, indeed, it appears
that this acceleration has been detected at about the magnitude
predicted in these models \citep{Riess:1998ab,
Perlmutter:1999ab}. Consistent with this, CMB anisotropy observations
are consistent with flat spatial hypersurfaces, which in conjunction
with the low mass-energy density parameter for non-relativistic matter
also requires a significant amount of dark energy.  These issues are
discussed in more detail in Sec.\ \ref{constraints} below and in
reviews \citep{Peebles:2003ab, Steinhardt:2003ab, Carroll:2004cd,
Padmanabhan:2005ab, Perivolaropoulos:2006ab, Copeland:2006ab,
Nobbenhuis:2006ab, Sahni:2006ab}.

The following sections flesh out this ``standard model" of cosmology,
elaborating on the model as well as describing the measurements and
observations on which it is based.

\section{Growth of Structure}
\label{growth}

\subsection{Gravitational Instability and Microphysics in the
          Expanding Universe}
\label{gravity}

\subsubsection{Gravitational Instability Theory from Newton Onwards}

The primary driver for the formation of large-scale structure in the
Universe is gravitational instability. The detailed growth of
structure depends on the nature of the initial fluctuations, the
background cosmology, and the constituents of the mass-energy density,
as causal physics influences the rate at which structure may grow on
different scales.

Newton, prompted by questions posed to him by Bentley, 
realized that a gas of randomly positioned
massive particles interacting gravitationally in flat spacetime is
unstable, and that as time progresses the mass density distribution
grows increasingly more anisotropic and inhomogeneous. Awareness of
this instability led Newton to abandon his preference for a finite and
bounded Universe of stars for one that is infinite and homogeneous
on average \citep[see discussion in][]{Harrison:2001ab}; this 
was an early discussion of the cosmological principle.

\citet{Jeans:1902ab} studied the stability of a spherical
distribution of gravitating gas particles in flat spacetime, motivated
by possible relevance to the process of star formation. He discovered that gas
pressure prevents gravitational collapse on small spatial scales and
gives rise to acoustic oscillations in the mass density inhomogeneity,
as the pressure gradient and gravitational forces compete. On large
scales the gravitational force dominates and mass density
inhomogeneities grow exponentially with time. The length scale on
which the two forces balance has come to be known as the Jeans length
or the acoustic Hubble length $c_s/H_0$, where $c_s$ is the speed of
sound.

On scales smaller than the Jeans length, adiabatic energy density
perturbations oscillate as acoustic waves. On scales well
below the Jeans length dissipative fluid effects (e.g., viscosity and
radiation diffusion) must be accounted for. These effects remove
energy from the acoustic waves, thus damping them. In an expanding
Universe, damping is effective when the dissipation time scale is 
shorter than the expansion time scale, and the smallest length scale 
for which this is the case is called the damping length. This is 
discussed in more detail below.

\subsubsection{Structure Growth in an Expanding Universe}

Study of gravitational instability in an evolving spacetime,
appropriate for the expanding Universe, began with Lema\^{\i}tre in
the early 1930's.  He pioneered two approaches, both of which are
still in use: a ``nonperturbative" approach based on a spherically
symmetric solution of the Einstein equations (further developed by
Dingle, Tolman, Bondi, and others and discussed in Sec.\ 
\ref{galformation} below); and a ``perturbative" approach in which 
one studies small departures from spatial homogeneity and isotropy 
evolving in homogeneous and isotropic background spacetimes.

At early times, and up to the present epoch on sufficiently large
scales, the growth of structure by gravitational instability is
accurately described by linear perturbation theory.  The growth of
small density and velocity perturbations must take into account the
effects of the expansion of the Universe. A fully 
relativistic theory must be employed to describe the growth of structure, 
because it is necessary to also describe the evolution of modes with 
wavelength larger than the Hubble length. In contrast, a Newtonian
approximation is valid and used on smaller length scales.

\citet{Lifshitz:1946ab} laid the foundations of
the general-relativistic perturbative approach to structure formation. 
He linearized the Einstein and stress-energy conservation equations about
a spatially homogeneous and isotropic Robertson-Walker background spacetime
metric and decomposed the departures from homogeneity and isotropy
into independently evolving spatial harmonics (the so-called scalar, 
vector, and tensor modes). Lifshitz treated matter as a fluid which
is a good approximation when the underlying particle mean free
path is small. He discovered that the vector transverse peculiar 
velocity (the peculiar velocity is the velocity that remains after 
subtracting off that due to the Hubble expansion) perturbation decays 
with time as a consequence of angular momentum conservation and 
that the contemporary Universe could contain a residual tensor 
gravitational wave background left over from earlier times.

Unlike the exponentially growing energy density irregularity that
Jeans found in flat spacetime on large scales, Lifshitz found only a
much slower power-law temporal growth, leading him to the incorrect
conclusion that ``gravitational instability is not the source of
condensation of matter into separate nebulae". It was almost two
decades before \citet{Novikov:1964ab} \citep[but see][for an earlier
hint]{Bonnor:1957ab} corrected this misunderstanding, noting that even
with power-law growth there was more than enough time for
inhomogeneities to grow, since they could do so even while they were
on scales larger than the Hubble length $r_H = c/H_0$ at early time.

The approach to the theory of linear perturbations initiated by
Lifshitz is based on a specific choice of spacetime coordinates
called synchronous coordinates. This approach is discussed in detail in
Sec.\ V (also see Sec.\ II) of \citet{Peebles:1980rh}, Sec.\ III of 
\citet{Zeldovich:1983ab}, \citet{Ratra:1988cd}, and other
standard cosmology and astroparticle physics
textbooks. \citet{Bardeen:1980ab} (building on 
earlier work) recast the Lifshitz analysis in a 
coordinate-independent form, and this approach has also become 
popular. For reviews of this approach see 
\citet{Mukhanov:1992ab}, as well as the standard textbooks.

A useful formalism for linear growth of density and velocity fields is
given by the ``Zel'dovich approximation" \citep{Zeldovich:1970cd,
Shandarin:1989ab, Sahni:1995ab}, based on anisotropic collapse and 
so ``pancake" formation (a concept earlier discussed in the context of the
initial singularity). This method accurately describes structure
formation up to the epoch when nonlinearities become significant.
Numerical simulations (see Sec.\ \ref{sims} below) of fully non-linear
structure growth often employ the Zel'dovich approximation for setting
the initial conditions of density and velocity.

\subsubsection{Space Curvature}

The evolution of the background spacetime influences the rate of
growth of structure. An early example of this effect is seen in the 
\citet{Gamow:1939ab} approximate generalization of 
Jeans' analysis to the expanding Universe, in particular to a model 
with open spatial hypersurfaces. At late times the dominant form of 
energy density in such a model is that due to the curvature of spatial 
hypersurfaces, because this redshifts away slower than the energy density 
in nonrelativistic matter.  The gravitational instability growth rate is
determined by the matter energy density, but the expansion rate
becomes dominated by the space curvature. As a result, the Universe expands
too fast for inhomogeneities to grow and large-scale structure
formation ceases. [A quarter century later, \citet{Peebles:1965ab} 
noted the importance of this effect.] This was the first example of an 
important and general phenomenon: a dominant spatially-homogeneous 
contributor to the energy density budget will prevent the growth of 
irregularity in matter.

\subsubsection{Dark Energy}

Matter perturbations also cannot grow when a cosmological constant or
nearly homogeneous dark energy dominates. There is strong evidence 
that dark energy --- perhaps in the form of Einstein's cosmological 
constant --- currently contributes $\sim 70$~\% of the mass-energy 
density of the universe. This dark energy was sub-dominant until 
recently, when it started slowing the rate of growth of structure
\citep{Peebles:1984ab}, thus its effect on dynamical evolution is 
milder than that of space curvature.

\subsubsection{Radiation and its Interaction with Baryonic Matter}

\citet{Guyot:1970ab} showed that a dominant homogeneous 
radiation background makes the Universe expand too fast to allow 
matter irregularities to start growing until the model becomes 
matter dominated (when the radiation redshifts away). Because of 
this effect, as discussed next, the acoustic Hubble length at the 
epoch when the densities of matter and radiation are equal is an 
important scale for structure formation in the expanding Universe. 
This imprints a feature in the power spectrum of matter fluctuations 
on the scale of the acoustic Hubble length at matter-radiation 
equality that can be used to measure the cosmic density of 
non-relativistic matter. We return to this in Sec.\ \ref{constraints}; 
a related CMB anisotropy effect is discussed in the next subsection 
\ref{anisotropy}.

\citet{Gamow:1948ab} noted that at early times in the 
Big Bang Model radiation (which has large relativistic pressure) 
dominates over baryonic matter.  In addition, at high temperature radiation 
and baryonic matter are strongly coupled by Thomson-Compton scattering 
and so behave like a single fluid. As a result of the large radiation 
pressure during this early epoch the Jeans or acoustic Hubble length 
is large and so gravitational growth of inhomogeneity occurs only on 
large scales, with acoustic oscillations on small scales. 
\citet{Peebles:1970ab} develop this picture.  

As the Universe cooled down below a temperature $T \sim 3000$ K at a 
redshift $z \sim 10^3$, the radiation and baryons decoupled.  Below 
this temperature proton nuclei can capture and retain free electrons 
to form electrically neutral
hydrogen atoms --- this process is called ``recombination'' --- because
fewer photons remained in the high energy tail of the distribution
with enough energy to disassociate the hydrogen atoms.  \citet{Peebles:1968ab}
and \citet{Zeldovich:1968ab} perform an
analysis of cosmological recombination, finding that at the ``end" of 
recombination there were enough charged particles left over for the 
Universe to remain a good conductor all the way to the present.  The 
finite time required for recombination results in a surface of non-zero 
thickness within which the decoupling of now-neutral baryons and photons 
occurs. The mean-free path for photons quickly grew, allowing the photons 
to travel (almost) freely, thus this ``last-scattering surface'' is the 
``initial'' source of the observed CMB; it is an electromagnetically opaque
``cosmic photosphere''.  See the standard cosmology textbooks for 
discussions of recombination.

Decoupling leads to a fairly steep drop in the pressure of the baryon
gas, and so a fairly steep decrease in the baryon Jeans length.
\citet{Peebles:1965ab} was developing this picture as the CMB was
being discovered. \citet{Peebles:1968cd} \citep[also
see][]{Peebles:1967ab} noted that the baryonic Jeans mass after
decoupling is of the order of the mass of a typical globular cluster
and so proposed that proto-globular-clusters were the first objects to
gravitationally condense out of the primordial gas. This model would
seem to predict the existence of extra-galactic globular clusters,
objects that have not yet been observationally recognized. There are,
however, dwarf galaxies of almost equally low mass, and we now also
know that some globular clusters are young and so globular clusters
might form in more than one way \citep[for a recent review
see][]{Brodie:2006ab}.

On scales smaller than the Jeans mass, dissipative 
effects become important and the ideal fluid approximation for 
radiation and baryonic matter is no longer accurate. As the Universe 
cools down towards recombination and decoupling, the photon mean 
free path grows and so photons diffuse out of more dense regions to 
less dense regions. As they diffuse the photons drag some of the 
baryons with them and so damp small-scale inhomogeneities in the 
photon-baryon fluid. This collisional damping --- a consequence of 
Thomson-Compton scattering --- is known as Silk damping in the cosmological 
context; it was first studied by \citet{Michie:1969ab}\footnote{
This is a version of a manuscript submitted to the Astrophysical
Journal on September 1, 1967, and only minimally revised (in response
to the referee's suggestions) before the author died.}, 
\citet{Peebles:1967ab}, and \citet{Silk:1968ab}. The Silk damping 
scale is roughly that of a cluster of galaxies.

\subsubsection{Possible Matter Constituents}


If baryons were the only form of non-relativistic matter the
density of matter would be so low that the Universe would remain
radiation dominated until after recombination.  The expansion rate
would be too large for gravitational instability to cause
inhomogeneity growth until matter starts to dominate well after last
scattering. The short time allowed for the gravitational growth of
inhomogeneity from the start of matter domination to today would 
require a large initial fluctuation amplitude to produce the observed 
large-scale structure. This scenario is ruled out by measurements of the
anisotropy of the CMB which indicate that fluctuations in the baryons
at decoupling are too small to have grown by gravitational instability
into the structures seen today in the galaxy distribution.


A solution to this puzzle is provided by dark matter, of the same type
and quantity needed to explain gravitational interactions on galactic
and cluster scales. Including this component of matter the Universe
becomes matter dominated at a redshift comparable to, or even larger
than, the redshift of last scattering.  Because CDM does not directly
couple to radiation, inhomogeneities in the distribution of CDM
begin to grow as soon as the Universe becomes matter dominated. Growth
in structure in the baryons, on scales small compared to the Hubble 
length, remains suppressed by Thomson-Compton scattering
until recombination, after which baryons begin to gravitate toward the
potential wells of dark matter and the baryon fluctuation amplitude
quickly grows. Thus, the low observed CMB anisotropy is reconciled
with observed large-scale structure.  (The CMB, while not directly
coupled to the CDM, feels the gravitational potential fluctuations of
the CDM.  Consequently, measurements of the CMB anisotropy probe the
CDM distribution.)  This is an independent, although model-dependent and
indirect, argument for the existence of CDM.


As mentioned above in Sec.\ \ref{nucleosynthesis}, the Universe also 
contains low mass neutrinos (precise masses are not yet known). These 
neutrinos are relativistic and weakly coupled (nearly collisionless) 
and so have a very long mean free path or free-streaming length. 
Consequently, they must be described by a distribution function, not a 
fluid. Because they are 
relativistic they have a large Jeans mass and gravitational instability 
is effective at collecting them only on very large scales, i.e., 
low mass neutrinos suppress power on small and intermediate length
scales. This effect makes it possible to observationally probe 
these particles with cosmological measurements \citep{Elgaroy:2005ab, 
Lesgourgues:2006ab}.

\subsubsection{Free Streaming}

Thus, the properties of dark matter are reflected in the spectrum of
density fluctuations because scales smaller than the free-streaming
scale of massive particles are damped \citep{Bond:1980ab}.  For hot
dark matter (HDM), e.g., neutrinos, the free-streaming scale is larger 
than the Hubble length at matter-radiation equality, hence the spectrum 
retains only large-scale power.  In such a ``top-down" 
scenario, superclusters form first, then fragment into smaller structures
including clusters of galaxies and individual galaxies, as first discussed by
Zel'dovich and collaborators.  The top-down model was inspired by 
experimental suggestions (now known to be incorrect) that massive neutrinos 
could comprise the nonbaryonic dark matter, and by an early (also now
known to be incorrect) interpretation of observational data on 
superclusters and voids (see Sec.\ \ref{spectro} below) that 
postulated that these were the basic organizational blocks for large-scale 
structure.  It predicts that smaller scale structure (e.g., 
galaxies) is younger than larger scale structure (e.g., superclusters), 
contrary to current observational indications. In fact, these observational 
constraints on the evolution of structure constrain the amount of HDM 
neutrino matter-energy density and so neutrino masses 
\citep{Kahniashvili:2005ab}. Cosmological observations provide the 
best (model-dependent) upper limits on neutrino masses.

For CDM, e.g., weakly interacting massive particles (WIMPs), the 
free-streaming scale is negligible for cosmological purposes.  
This ``bottom-up" or ``hierarchical" scenario, pioneered by Peebles and 
collaborators, begins with the formation
of bound objects on small scales that aggregate into larger structures,
thus galaxies result from mergers of sub-galaxies, with superclusters
being the latest structures to form. This is in better agreement with the
observational data. See Sec.\ \ref{dark} for more details on this model.

\subsubsection{Initial Density Perturbations and the Transfer Function}

The current standard model for structure formation assumes that
structure in the Universe arose primarily from gravitational
amplification of infinitesimal scalar density perturbations in the
early Universe. The processes listed in this section modify these
initial inhomogeneities. Reviews are given in \citet[Sec.\
V]{Peebles:1980rh}, \citet[Sec.\ III]{Zeldovich:1983ab},
\citet{Efstathiou:1990ab}, \citet[Ch.\ 9]{Kolb:1990ab}, \citet[Ch.\
4]{Padmanabhan:1993ab}, \citet{Dekel:1999ab}, and \citet[Part
II]{Mukhanov:2005ab}.

As discussed in Secs.\ \ref{dark} and \ref{anisotropy}, observations 
to date are consistent with primordial fluctuations that are Gaussian 
random phase.  These are the type of fluctuations expected if the seeds
for structure formation result from the superposition of quantum
mechanical zero-point fluctuations of the scalar field that drove
inflation of the early Universe, in the simplest inflation models, as
discussed in Sec.\ \ref{inflation} above. In the simplest inflation models the
fluctuations are adiabatic. Furthermore, observational data are
consistent with only adiabatic perturbations, so in what follows we
focus on this case [see \citet{Bean:2006ab} for a recent
discussion of constraints on isocurvature models].

As discussed in Sec.\ \ref{inflation} above and Sec.\ \ref{constraints}
below, current large-scale observational results are reasonably well
fit by an $n=1$ scale-invariant primordial spectrum of perturbations,
the kind considered by \citet{Harrison:1970ab}, \citet{Peebles:1970ab}, 
and \citet{Zeldovich:1972ab}, and
predicted in some of the simpler inflation models.  The effect of
causal physics on the later growth of structure, as discussed above,
may then be represented by a ``transfer function'' that describes the
relative growth of fluctuations on different wavelength
scales. Observations of the anisotropy of the CMB and the clustering
of galaxies and clusters at the present epoch probe the shape of the
transfer function (as well as the primordial spectrum of
perturbations) and thereby constrain structure formation models.  Such
observations are discussed below in Secs.\ \ref{anisotropy} and \ref{mapping}.

\subsubsection{Gravitational Waves and Magnetic Fields}

As noted in Sec.\ \ref{inflation} above, more complicated models of
inflation can generate gravitational wave or magnetic field
fluctuations that break adiabaticity. A primordial magnetic field
might provide a way of explaining the origin of the uniform part of
contemporary galactic magnetic fields; there are enough charged
particles left over after recombination to ensure that primordial
magnetic field lines will be pulled in, and the field amplified, by a
collapsing gas cloud. \citet{Maggiore:2000ab} and \citet{Buonanno:2004ab} 
review primordial gravity waves, and cosmological magnetic fields are 
reviewed by \citet{Widrow:2002ab} and \citet{Giovannini:2004ab}. In the 
next subsection we consider the effects of such fields on the CMB.

\subsection{CMB Anisotropies}
\label{anisotropy}

As a result of the gravitational growth of inhomogeneities in the
matter distribution, when the photons decouple from the baryons at
last scattering at a redshift $z \sim 10^3$ (see Sec.\ \ref{gravity} 
above) the
photon temperature distribution is spatially anisotropic. In addition, in the
presence of a CMB temperature quadrupole anisotropy, Thomson-Compton
scattering of CMB photons off electrons prior to decoupling generates 
a linear polarization anisotropy of the CMB. After decoupling the CMB 
photons propagate almost freely, influenced only by gravitational 
perturbations and late-time reionization.  Measurements of the 
temperature anisotropy and polarization anisotropy provide important 
constraints on many parameters of models of structure formation. This 
area of research has seen spectacular growth in the last decade or so, 
following the COBE discovery of the CMB temperature anisotropy. It has 
been the subject of recent reviews; see \citet{White:2002ab},
\cite{Hu:2002ab}, \citet[Sec.\ IV.B.11]{Peebles:2003ab}, 
\citet{Subramanian:2005ab}, \citet{Giovannini:2005ab}, and 
\citet{Challinor:2005ab}. Here we focus only on a few recent developments.

The three-year WMAP observations of CMB temperature anisotropies
\citep{Hinshaw:2006ab} are state-of-the-art data.  On all but the very
largest angular scales, the WMAP data are consistent with the
assumption that the CMB temperature anisotropy is well-described by a
spatial Gaussian random process \citep{Komatsu:2003ab}, consistent with
earlier indications \citep{Park:2001ab, Wu:2001ab}.  The few
largest-scale angular modes exhibit a lack of power compared to what
is expected in a spatially-flat CDM model dominated by a cosmological
constant \citep{Bennett:2003ab}, resulting in some debate about the
assumptions of large-scale Gaussianity and spatial isotropy. 
This feature was also seen in the COBE data \citep{Gorski:1998ab}.
The estimated large-angular-scale CMB temperature anisotropy power depends
on the model used to remove foreground Galactic emission
contamination.  Much work has been devoted to understanding
foreground emission on all scales \citep[e.g.,][]{Mukherjee:2003ab,
Bennett:2003cd, Tegmark:2003ab}, and the current consensus is that
foregrounds are not the cause of the large-angular-scale WMAP effects. 

The CMB temperature anisotropy is conventionally expressed as an
expansion in spherical harmonic multipoles on the sky, and for a
Gaussian random process the multipole (or angular) power spectrum
completely characterizes the CMB temperature anisotropy. The observed
CMB anisotropy is reasonably well fit by assuming only adiabatic
fluctuations with a scale-invariant power spectrum. These
observational results are consistent with the predictions of the
simplest inflation models, where quantum-mechanical fluctuations in a
weakly-coupled scalar field are the adiabatic, Gaussian seeds for the
observed CMB anisotropy and large-scale structure.

Smaller-scale inhomogeneities in the coupled baryon-radiation fluid
oscillate (see Sec.\ \ref{gravity} above), and at decoupling some of 
these modes
will be at a maximum or at a minimum, giving rise to acoustic peaks
and valleys in the CMB anisotropy angular spectrum.  The relevant 
length scale is the acoustic Hubble length at the epoch of recombination;
this may be predicted by linear physics and so provides a standard ruler 
on the sky. Through the angular diameter distance relation, the multipole
numbers ${\ell}$ of oscillatory features in the temperature anisotropy
spectrum $C_{\ell}$ reflect space curvature ($\Omega_{\rm K}$) and the 
expansion history (which depends on $\Omega_{\rm M}$ and $\Omega_{\Lambda}$) 
of the Universe.  The angular scales of the peaks are sensitive to the
value of the matter density parameter in an open Universe, but not in
a spatially-flat ($\Omega_{\rm K} = 0$) Universe dominated by a 
cosmological constant, where the first peak 
is at a multipole index ${\ell} \sim 220$.  This provides a useful way 
to measure the curvature of spatial hypersurfaces. \citet{Sugiyama:1992ab}
and \citet{Kamionkowski:1994ab, Kamionkowski:1994cd} are 
early discussions of the CMB temperature anisotropy in an open Universe, 
and \citet{Brax:2000ab}, \citet{Baccigalupi:2002ab}, \citet{Caldwell:2004ab},
and \citet{Mukherjee:2003cd} consider the case of scalar field dark energy 
in a spatially-flat Universe. CMB temperature anisotropy data on the
position of the first peak is consistent with flat spatial
hypersurfaces \citep[e.g.,][]{Podariu:2001ab, Durrer:2003ab,
Page:2003ab}. Model-based CMB data analysis is used to constrain more
cosmological parameters \citep[e.g.,][]{Lewis:2002ab,
Mukherjee:2003ef, Spergel:2006ab}.  For example, the relative
amplitudes of peaks in this spectrum are sensitive to the mass
densities of the different possible constituents of matter (e.g., CDM,
baryons, and neutrinos, $\Omega_{\rm CDM}, \Omega_{\rm B}$, and 
$\Omega_{\nu}$).  

The CMB polarization anisotropy was first detected from the ground
by the DASI experiment at the South Pole \citep{Kovac:2002ab}.
The three-year WMAP observations are the current state of the art
\citep{Page:2006ab}. For a recent review of polarization measurements
see \citet{Balbi:2006ab}. The polarization anisotropy peaks at
a larger angular scale than the temperature anisotropy, indicating that 
there are inhomogeneities on scales larger than the acoustic Hubble
length at recombination, consistent with what is expected in the 
inflation scenario. The polarization anisotropy signal is interpreted as the 
signature of reionization of the Universe. The ability of WMAP to measure
polarization anisotropies allows this experiment to probe the early
epochs of non-linear structure formation, through sensitivity to the
reionization optical depth $\tau$.  

Primordial gravitational waves or a primordial magnetic field can also
generate CMB anisotropies. Of particular current interest are their
contributions to various CMB polarization anisotropies.  (Because
polarization is caused by quadrupole fluctuations, these anisotropies
constrain properties of the primordial fluctuations, such as the
ratio of tensor to scalar fluctuations, $r$.)  The effects of gravity
waves on the CMB are discussed in the more recent standard cosmology and 
astroparticle textbooks and by
\citet{Giovannini:2005ab}. The magnetic field case is reviewed by
\citet{Giovannini:2006ab} and \citet{Subramanian:2006ab}; recent topics of
interest may be traced from \citet{Lewis:2004ab}, \citet{Kahniashvili:2005cd,
Kahniashvili:2007ab}, and  
\citet{Brown:2005ab}.

We continue discussion of the CMB anisotropies and cosmological 
parameters in Sec.\ \ref{constraints}.

\subsection{Galaxy Formation and the End of the Dark Age}
\label{galformation}

The emission of the first light in the Universe, seen today as the
CMB, is followed by a ``dark age" before the first stars and quasars
form. \citet{Bromm:2004ab} review formation of the first stars.
Eventually, high energy photons from stars and quasars reionize
intergalactic gas throughout the Universe \citep[for reviews see][]{Fan:2006ab,
Choudhury:2006ab, Loeb:2006ab, Loeb:2006cd}. Observations
of polarization of microwave background photons by WMAP
\citep{Page:2006ab} suggest that reionization occurs at redshift
$z\approx 11$. However, strong absorption of Lyman-$\alpha$ photons by
intergalactic neutral hydrogen \citep{Gunn:1965ab}, seen in spectra of
quasars at redshift $z\approx 6$ \citep{Becker:2001ab, Fan:2002ab}
indicates that reionization was not complete until somewhat later.
This is an area of ongoing research \citep[see, e.g.,][]{Choudhury:2006cd,
Gnedin:2006ab, Alvarez:2006ab}.

Current models for galaxy formation follow the picture
\citep{Hoyle:1953ab, Silk:1977ab, Binney:1977ab, Rees:1977ab,
White:1978ab} in which dark matter halos form by collisionless
collapse, after which baryons fall into these potential wells, are
heated to virial temperature, and then cool and condense at the centers of
the halos to form galaxies as we know them.  In short, baryons fall
into the gravitational potentials of ``halos'' of dark matter, at the
same time that those halos grow in size, hierarchically aggregating
small clumps into larger ones. The baryons cool by emitting radiation 
and shed angular momentum, leading to concentrations of star formation and
accretion onto supermassive black holes within the dark matter halos.

In addition to the perturbative approach to structure formation
discussed in Sec.\ \ref{gravity}, Lema\^{\i}tre also pioneered a
``nonperturbative" approach based on a spherically symmetric solution
of the Einstein equations. This spherical accretion model
\citep{Gunn:1972ab} describes the salient features of the growth of
mass concentrations.  See \citet{Gott:1977ab}, \citet[Sec.\
22]{Peebles:1993ab}, and \citet{Sahni:1995ab} for reviews of such
models.

A phenomenological prescription for the statistics of non-linear 
collapse of structure, i.e., the formation of gravitationally bound 
objects, is given by the Press-Schechter formulae \citep{Press:1974ab, 
Sheth:1999ab}. Attempts to firm up the theoretical basis of such
formulae form the ``excursion set'' formalism which treats the formation 
of a gravitationally bound halo as the result of a random walk 
\citep{Mo:1996ab, White:1996ab, Sheth:2001ab}. For a review see
\citet{Cooray:2002ab}. These methods provide probability distributions 
for the number of bound objects as a function of mass threshold, and 
can be generalized to develop a complementary description of the 
evolution of voids \citep{Sheth:2004ab}. A more rigorous approach 
assumes structure forms at high peaks in the smoothed density field 
\citep{Kaiser:1984ab, Bardeen:1986ab, Sahni:1995ab}. Recent reviews of 
galaxy formation include \citet{Avila-Reese:2006ab} and \citet{Baugh:2006ab}. 
The next subsection, \ref{sims}, describes numerical methods for studying 
structure formation.

Apparent confirmation of the hierarchical picture of structure
formation includes the striking images of galaxies apparently in the
process of assembly obtained by the HST in the celebrated ``Hubble
Deep Fields" \citep{Ferguson:2000ab, Beckwith:2006ab}.  The detailed
properties of galaxies and their evolution are outside the scope of
this review.  Recent reviews of the observational situation are
\citet{Gawiser:2006ab} and \citet{Ellis:2007ab}. Texts covering this
topic include \citet{Spinrad:2005ab} and \citet{Longair:2008ab} .

While the current best model of structure formation, in which CDM
dominates the matter density, works quite well on large scales,
current observations indicate some possible problems with the CDM
model on smaller scales; see \citet{Tasitsiomi:2003ab}, \citet[Sec.\
IV.A.2]{Peebles:2003ab}, and \citet{Primack:2005ab} for reviews.
Simulations of structure formation indicate that CDM model halos may
have cores that are cuspier \citep{Navarro:1997ab, Swaters:2000ab} and
central densities that are higher \citep{Moore:1999ab, Firmani:2001ab}
than are observed in galaxies. Another concern is that CDM models
predict a larger than observed number of low-mass satellites of
massive galaxies \citep{Moore:1999cd, Klypin:1999ab}.  These issues
have led to consideration of models with reduced small-scale power.
However, it seems difficult to reconcile suppression of small-scale
power with the observed small-scale clustering in the neutral hydrogen
at redshifts near 3.

The relationship between the distributions of galaxies (light) and 
matter is commonly referred to as ``biasing." The currently-favored 
dark energy dominated CDM model does not require significant bias 
between galaxies and matter; in the best-fit model the ratio of 
galaxy to matter clustering is close to unity for ordinary galaxies 
\citep{Tegmark:2004cd}.

\subsection{Simulations of Structure Formation}
\label{sims}

Cosmological simulations using increasingly sophisticated numerical
methods provide a testbed for models of structure formation.
\citet{Bertschinger:1998ab} reviews methods and results.

Computer simulations of structure formation in the Universe began with
purely gravitational codes that directly compute the forces between a
finite number of particles (``Particle-Particle" or PP codes) that sample 
the matter distribution. Early results used direct $N$-body calculations
\citep{Aarseth:1979ab}. Binning the particles on a grid and computing the 
forces using the Fast Fourier Transform (the ``Particle-Mesh" or PM method) 
is computationally more efficient, allowing simulation of larger volumes
of space, but has force resolution of the order of the grid spacing. A
compromise is the P$^3$M method, which uses PM for large scale forces
supplemented by direct PP calculations on small scales, as used for
the important suite of CDM simulations by \citet{Davis:1985ab}. 
For details on these methods see \citet{Hockney:1988ab}.

The force resolution of PM codes and the force resolution and speed 
of P$^3$M codes may be increased by employing multiple grid levels 
\citep{Villumsen:1989ab, Couchman:1991ab, Bertschinger:1991ab, Gnedin:1996ab}. 
Adaptive mesh refinement \citep[AMR;][]{Berger:1989ab} does this dynamically to
increase force resolution in the PM gravity solver \citep{Kravtsov:1997ab, 
Norman:1999ab}. 

Another approach to achieving both speed and good force resolution in
gravitational $N$-body simulation is use of the hierarchical tree
algorithm \citep{Barnes:1986ab}. Large cosmological simulations have
used a parallelized version of this method \citep{Zurek:1994ab}. 
Significant increase in speed was found with the Tree Particle-Mesh 
algorithm \citep{Bode:2000ab}.  GOTPM \citep{Dubinski:2004ab}, a 
parallelized hybrid PM+tree scheme, has been used for simulations 
involving up to $8.6\times 10^9$ particles. PMFAST \citep{Merz:2005ab} 
is a recent parallelized multi-level PM code.

Incorporation of hydrodynamics and radiative transfer in cosmological
simulations has made it possible to study not only the gravitational
formation of dark matter halos, but also the properties of baryonic
matter, and thus the formation of galaxies associated with those
halos.  Methods for solving the fluid equations include
smooth-particle hydrodynamics [SPH; see \citet{Monaghan:1992ab} for a
review], which is an inherently Lagrangian approach,
and Eulerian grid methods. Cosmological SPH simulations were pioneered
by \citet{Evrard:1988ab} and \citet{Hernquist:1989ab}.  To date, the
cosmological simulation with the largest number of particles
($10^{10}$) employs SPH and a tree algorithm
\citep[GADGET;][]{Springel:2001ab}.  Grid-based codes used for
cosmological simulation include that described by \citet{Cen:1992ab}
and \citet{Ryu:1993ab}.

To date, no code has sufficient dynamic range to compute both the
large scale cosmological evolution on scales of many hundreds of
megaparsecs and the formation of stars from baryons, but physical
heuristics have been successfully incorporated into some codes to
model the conversion of baryons to stars \citep[see, e.g.,][]{Cen:1992ab}.

The Millenium Run simulation \citep{Springel:2005ab} represents the
current state-of-the art in following the evolution of both the dark
matter and baryonic components on scales from the box size,
$500h^{-1}$ Mpc, down to the resolution limit of roughly
$5h^{-1}$ kpc. See this article and references therein for discussion
of the many pieces of uncertain physics necessary for producing the
observed baryonic structures.

Another approach to modeling the properties of the galaxies
associated with dark matter halos is to use the history of halo
mergers together with semi-analytic modeling of galaxy properties
\citep{Lacey:1993ab, Kauffmann:1993ab, Cole:1994ab, Somerville:1999ab}.
When normalized to the observed luminosity function of galaxies and 
Tully-Fisher relation for spiral galaxies, these semi-analytic models 
(SAMs) reproduce many of the observed features of the galaxy 
distribution. A common approach is to use SAMs to ``paint on" the 
properties of galaxies that would reside in the dark matter halos 
found in purely gravitational simulations. See \citet{Avila-Reese:2006ab}
and \citet{Baugh:2006ab} for recent reviews. Related to the SAMs 
approach are halo occupation models \citep{Berlind:2002ab, Kravtsov:2004ab} 
that parameterize the statistical relationship between the masses of dark 
matter halos and the number of galaxies resident in each halo.

\section{Mapping the Universe}
\label{mapping}

The observed features of the large-scale distribution of matter
include clusters, superclusters, filaments, and voids.  By mapping the
distribution of galaxies in the Universe, both in two dimensions as
projected on the sky and in three dimensions using spectroscopic
redshifts, astronomers seek to quantify these inhomogeneities in order
to test models for the formation of structure in the Universe.  Not
only the spatial distribution of galaxies, but also the distribution
of clusters of galaxies, quasars, and absorption line systems provide
constraints on these models. Peculiar velocities of galaxies, which
reflect inhomogeneities in the mass distribution, provide further
constraints. Here we briefly review important milestones and surveys
relevant for testing cosmological models.

\subsection{Galaxy Photometric Surveys}
\label{photo}

Studies of the global spacetime of the Universe assume the 
``cosmological principle'' \citep{Milne:1933ab} which is the
supposition that the Universe is statistically homogeneous when viewed on
sufficiently large scales.  The angular distribution of radio galaxies
provides a good test of this approach to homogeneity, because
radio-bright galaxies and quasars may be seen in flux-limited samples
to nearly a Hubble distance, $c/H_0$.  Indeed, the $\sim 31,000$ 
brightest radio galaxies observed at a wavelength of 6 cm 
\citep{Gregory:1991ab} are distributed nearly isotropically, and similar 
results are found in the FIRST radio survey \citep{Becker:1995ab}.  
[For a review of other evidence for large-scale spatial isotropy see 
Sec.\ 3 of \citet{Peebles:1993ab}.] In contrast, the Universe is 
clearly inhomogeneous on the more modest scales probed by optically-selected
samples of bright galaxies, For example, significant clustering is
observed among the roughly $30,000$ galaxies in the \citet{Zwicky:1961ab} 
catalog.

Maps of the distribution of nebulae revealed anisotropy in the sky
before astronomers came to agree that many of these nebulae were
distant galaxies \citep{Charlier:1925ab}.  The \citet{Shapley:1932ab}
catalog of galaxies clearly showed the nearby Virgo 
cluster of galaxies. Surveys of selected areas on the sky using 
photographic plates to detect distant galaxies clearly revealed 
anisotropy of the galaxy distribution and were used to quantify 
this anisotropy \citep{Mowbray:1938ab}. \citet{de Vaucouleurs:1953ab} 
recognized in this anisotropy the projected 
distribution of the local supercluster of galaxies.

\citet{Rubin:1954ab} used two-point correlations of galaxy counts
from Harvard College Observatory plates to detect fluctuations on the
scale of clusters of galaxies. The \citet{Shane:1954ab} Lick Survey of 
galaxies used counts of galaxies found on large-format 
photographic plates taken at Lick Observatory to make the first large-scale 
map of the angular distribution of galaxies suitable for statistical 
analysis.  Early analysis of these data included methods such as 
counts-in-cell analyses and the two-point correlation function 
\citep{Limber:1954ab, Totsuji:1969ab}. The sky map of the Lick counts
produced by \citet{Seldner:1977ab} visually demonstrated 
the rich structure in the galaxy distribution.  Peebles and collaborators 
used these data for much of their extensive work on galaxy clustering 
\citep{Groth:1977ab}; for a review see \citet[Sec.\ III]{Peebles:1980rh}.

The first Palomar Observatory Sky Survey (POSS) yielded two important
catalogs: the \citet{Abell:1958ab} catalog of clusters and the
\citet{Zwicky:1961ab} catalog of  clusters and galaxies identified by eye 
from the photographic plates.  \citet{Abell:1961ab} found
evidence for angular ``superclustering" (clustering of galaxy clusters)
that was confirmed statistically by \citet{Hauser:1973ab}.
Photographic plates taken at the UK Schmidt telescope were digitized
using the Automatic Plate Measuring (APM) machine to produce the APM
catalog of roughly two million galaxies. Calibration with CCD photometry 
made the APM catalog invaluable for
accurate study of the angular correlation function of galaxies on
large scales \citep{Maddox:1990ab}. Perhaps the last large-area galaxy 
photometric survey to employ photographic plates is the Digitized Palomar 
Observatory Sky Survey (DPOSS) \citep{Gal:2004ab}.

The largest imaging survey that employs a camera with arrays of
charge-coupled devices (CCDs) is the Sloan Digital Sky Survey 
\citep[SDSS;][]{Stoughton:2002ab}. The imaging portion of this survey includes
five-color digital photometry of $8000$ deg$^2$ of sky, with 215
million detected objects.  Imaging for the SDSS is obtained using a
special-purpose 2.5 m telescope with a three-degree field of view
\citep{Gunn:2006ab}.

Important complements to optical surveys include large-area catalogs
of galaxies selected in the infrared and ultraviolet.  Nearly all-sky
source catalogs were produced from infrared photometry obtained with
the Infrared Astronomical Satellite \citep[IRAS;][]{Beichman:1988ab} and
the ground-based Two Micron All Sky Survey 
\citep[2MASS;][]{Jarrett:2000ab}.  The ongoing Galaxy Evolution Explorer
satellite \citep[GALEX;][]{Martin:2005ab} is obtaining ultraviolet
imaging over the whole sky.

\subsection{Galaxy Spectroscopic Surveys}
\label{spectro}

Systematic surveys of galaxies using spectroscopic redshifts to infer
their distances began with observations of galaxies selected from the
Shapley-Ames catalog \citep{Humason:1956ab, Sandage:1978ab}.  Important
early mapping efforts include identification of superclusters and voids 
in the distribution of galaxies and Abell clusters by \citet{Joeveer:1978ab}, 
the \citet{Gregory:1978ab} study of the Coma/Abell1367 supercluster and 
its environs that identified voids, and the \citet{Kirshner:1981ab} 
study of the correlation function of galaxies and discovery of the giant 
void in Bo\"otes. 
Early targeted surveys include the 
\citet{Giovanelli:1985ab} survey of the Perseus-Pisces supercluster.

Redshift surveys of large areas of the sky began with the first Center
for Astrophysics redshift survey \citep[CfA1;][]{Huchra:1983ab}, which
includes redshifts for 2401 galaxies brighter than apparent magnitude
$m_B=14.5$ over a wide area toward the North Galactic Pole.  CfA2
\citep{Falco:1999ab} followed over roughly the same area, extending to
apparent magnitude $m_B=15.5$. At this depth, the rich pattern of
voids, clusters, and superclusters were strikingly obvious
\citep{deLapparent:1986ab}.  \citet{Giovanelli:1991ab} review the
status of galaxy redshift surveys ca.\ 1991.

Both CfA redshift surveys used the Zwicky catalog of galaxies to select 
targets for spectroscopy.  The Southern Sky Redshift Survey 
\citep[SSRS;][]{daCosta:1998ab} covers a large area
of the southern hemisphere (contiguous with CfA2 in the northern
galactic cap) to similar depth, using the ESO/Uppsala Survey to select
galaxy targets and a spectrograph similar to that employed for the CfA
surveys. The Optical Redshift Survey (ORS) supplemented existing
redshift catalogs with 1300 new spectroscopic redshifts to create a
nearly all-sky survey \citep{Santiago:1995ab}.

Deep ``pencil-beam" surveys of narrow patches on the sky revealed
apparently-periodic structure in the galaxy distribution 
\citep{Broadhurst:1990ab}.

The Las Campanas Redshift Survey \citep[LCRS;][]{Shectman:1996ab}, the first
large-area survey to use fiber optics, covered over 700 deg$^2$ near
the South Galactic Pole. This survey was important because it showed
that structures such as voids and superclusters found in shallower
surveys are ubiquitous but the structures seen by LCRS were no larger
than those found earlier.  The Century Survey \citep{Geller:1997ab} and
the ESO Deep Slice survey \citep{Vettolani:1998ab} were likewise useful
for statistically confirming this emerging picture of large-scale
structure.

Sparse surveys of galaxies to efficiently study statistical properties
of the galaxy distribution include the Stromlo-APM redshift survey
\citep{Loveday:1996ab} based on $1/20$ sampling of the APM galaxy
catalog and the Durham/UK Schmidt redshift survey \citep{Ratcliffe:1998ab}.

While optically-selected surveys are relatively blind to structure
behind the Milky Way, redshift surveys based on objects detected in
the infrared provide nearly all-sky coverage.  A sequence of surveys
of objects detected by IRAS were carried out, flux-limited to 2 Jy 
\citep{Strauss:1992ab}, 1.2 Jy \citep{Fisher:1995ab}, and 0.6 Jy 
\citep{Saunders:2000ab}. The 6dF Galaxy Survey \citep{Jones:2004ab} will 
measure redshifts of $150,000$ galaxies photometrically identified by 
2MASS \citep{Jarrett:2000ab}.

The 2-degree Field Galaxy Redshift Survey (2dFGRS) of $250,000$ galaxies
\citep{Colless:2001ab} was selected from the APM galaxy catalog and
observed using the two-degree field multi-fiber spectrograph at the
Anglo-Australian 4 m telescope. The survey is complete to apparent
magnitude $m_J=19.45$ and covers about 1500 deg$^2$.

The spectroscopic component of the SDSS \citep{Stoughton:2002ab}
includes medium-resolution spectroscopy of $675,000$ galaxies and
$96,000$ quasars identified from SDSS photometry. These spectra are
obtained with dual fiber-optic CCD spectrographs on the same 2.5 m
telescope.  The main galaxy redshift survey is complete to $m_r=17.77$
and is complemented by a deeper survey of luminous red galaxies. The
ongoing extension of this survey (SDSS-II) will expand the
spectroscopic samples to more than $900,000$ galaxies and $128,000$
quasars.

Spectroscopic surveys that trace structure in the galaxy distribution
at much larger redshift include the DEEP2 survey \citep{Coil:2004ab}
and others \citep{Steidel:2004ab} employing the Keck Observatory,
and the VIMOS VLT Deep survey \citep{LeFevre:2005ab}.

\subsection{Cluster Surveys}
\label{clusters}

Mapping of the Universe using galaxy clusters as tracers began with
study of the Abell catalog \citep{Abell:1958ab, Abell:1989ab}.
Studies of the angular clustering of Abell clusters includes
\citet{Hauser:1973ab}.  Several redshift surveys of Abell clusters
have been conducted, including those described by
\citet{Postman:1992ab} and \citet{Katgert:1996ab}.  Important cluster samples
have also been identified from digitized photographic plates from the
UK Schmidt telescope, followed up by redshift surveys of cluster
galaxies \citep{Lumsden:1992ab, Dalton:1992ab}.  More distant samples
of clusters have been identified using deep CCD photometry 
\citep[see, e.g.,][]{Postman:1996ab, Gladders:2005ab}.  In X-ray
bandpasses, cluster samples useful for studying large-scale structure
have been identified using data from ROSAT \citep{Romer:1994ab,
Boehringer:2004ab}.  The SDSS is producing large catalogs of galaxy
clusters using a variety of selection methods
\citep{Bahcall:2003ab}. Use of the Sunyaev-Zel'dovich effect (the
microwave decrement caused by Thomson-Compton scattering of the CMB
photons by the intracluster gas) holds great promise to identify new
deep samples of galaxy clusters \citep{Carlstrom:2002ab}.  General
reviews of clusters of galaxies include \citet{Rosati:2002ab},
\citet{Voit:2005ab}, and \citet{Borgani:2006ab}.

\subsection{Quasar surveys}
\label{quasars}

The advent of multi-object wide-field spectrographs has made possible
collection of very large samples of spectroscopically-confirmed
quasars, as observed by the 2dF QSO Redshift Survey
\citep{Croom:2004ab} and the SDSS \citep{Schneider:2005ab}.  For a ca.\
1990 review of the field see \citet{Hartwick:1990ab}.  While
quasars themselves are too sparsely distributed to provide good maps
of the large-scale distribution of matter, their clustering in
redshift space has been measured \citep{Osmer:1981ab} and generally
found to be similar to that of galaxies \citep{Outram:2003ab}.  Similar
results obtain from clustering analyses of active galactic nuclei in
the nearby universe \citep{Wake:2004ab}, although this clustering
depends in detail on the type of AGN \citep{Constantin:2006ab}.

The distribution of absorption lines from gas, particularly from the
Lyman-$\alpha$ ``forest" of neutral hydrogen clouds along the line of
sight toward bright quasars \citep{Lynds:1971ab, Rauch:1998ab} provides 
an important statistical probe of the distribution of matter 
\citep[see, e.g.,][]{McDonald:2005ab} on small scales and at large redshift.

\subsection{Peculiar Velocity Surveys}
\label{peculiar}

When measured over sufficiently large scales, the peculiar motions of
galaxies or clusters simply depend on the potential field generated by
the mass distribution \citep[see][]{Peebles:1980rh, Peebles:1993ab, 
Davis:1983ab}.  Techniques for measuring distances to other galaxies
are critically reviewed in \citet{RowanRobinson:1985ab},
\citet{Jacoby:1992ab}, \citet{Strauss:1995ab}, and
\citet{Webb:1999ab}. Together with the galaxy or cluster redshifts,
these measurements yield maps of the line-of-sight component of the
peculiar velocity. From such data it is possible to reconstruct a map
of the matter density field \citep[e.g.,][]{Bertschinger:1989ab,
Dekel:1994ab} or to trace the galaxy orbits back in time
\citep[e.g.,][]{Peebles:1990ab, Goldberg:2000ab}. Analyses of correlations of
the density and velocity fields also yield constraints on the cosmic
matter density \citep[e.g.,][]{Willick:1997ab}.

\citet{Rubin:1976ab} were the first to find evidence for
bulk flows from galaxy peculiar velocities. 
\citet{Dressler:1987cd} found evidence for a bulk flow toward a large
mass concentration, dubbed the ``Great Attractor."
\citet{Lauer:1994ab} found surprising evidence for motion of the Local
Group on a larger scale. However, analysis of subsequent peculiar
velocity surveys indicates that the inferred bulk flow results,
including those of Lauer and Postman, are consistent within the
uncertainties \citep{Hudson:2000ab}.  The status of this field ca.\
1999 is surveyed by \citet{Courteau:2000ab}, recent results
include \citet{Hudson:2004ab}, and \citet{Dekel:1994ab} and 
\citet{Strauss:1995ab} review this topic.  Comparison of peculiar
velocity surveys with the peculiar velocity of our Galaxy implied by
the CMB dipole indicates that a significant component of our motion
must arise from density inhomogeneities that lie at rather large
distance, beyond $60h^{-1}$ Mpc \citep{Hudson:2004ab}.

\subsection{Statistics of Large-Scale Structure}
\label{statistics}

The clustering pattern of extragalactic objects reflects both the
initial conditions for structure formation and the complex
astrophysics of formation and evolution of these objects. In the
standard picture described above, linear perturbation theory
accurately describes the early evolution of structure, thus
measurement of clustering on very large scales, where the clustering
remains weak, closely reflects the initial conditions. On these scales
the density field is very nearly Gaussian random phase, therefore the
two-point correlation function of the galaxy number density field
(also called the autocorrelation or covariance function) or its
Fourier transform, the power spectrum, provides a complete statistical
description. (Temperature anisotropies of the CMB discussed in
Sec. \ref{anisotropy} arise from density fluctuations at redshift
$z\sim 10^3$ that evolve in the fully linear regime.)  On the scales
of galaxies and clusters of galaxies, gravitational evolution is
highly non-linear and the apparent clustering depends strongly on the
detailed relationship between mass and light in galaxies.  In between
the linear and non-linear regimes lies the ``quasi-linear" regime in
which clustering growth proceeds most rapidly.  A wide range of
statistical methods have been developed to quantify this complex
behavior.  Statistical properties of the galaxy distribution and
details of estimating most of the relevant statistics are described in
depth by \citet{Peebles:1980rh}, \citet{Martinez:2002ab}, and
\citet{Bernardeau:2002ab}.  Methods of using galaxy redshift surveys to
constrain cosmology are reviewed by \citet{Lahav:2004ab} and 
\citet{Percival:2006ab}. Constraints on cosmological parameters from such
measurements are discussed below in Sec.\ \ref{constraints}.

The simplest set of statistical measures are the $n$-point correlation
functions, which describe the joint probability in excess of random of
finding $n$ galaxies at specified relative separation.  Early
applications of correlation functions to galaxy data include
\citet{Limber:1954ab}, \citet{Totsuji:1969ab}, and
\citet{Groth:1977ab}. The $n$-point functions may be estimated by
directly examining the positions of $n$-tuples of galaxies or by using
moments of galaxy counts in cells of varying size. Tests of scaling
relations among the $n$-point functions are discussed in detail by
\citet{Bernardeau:2002ab}.

Power spectrum analyses of large galaxy redshift surveys
\citep{Vogeley:1992ab, Fisher:1993ab, Tegmark:2004ab} yield useful
constraints on cosmological models. Closely
related to power spectrum analyses are estimates of cosmological
parameters using orthogonal functions \citep{Vogeley:1996ab, Pope:2004ab}.
\citet{Tegmark:1998ab} discuss the merits of different methods
of power spectrum estimation.  \citet{Verde:2002ab} describe a
measurement of the galaxy bispectrum.

A number of statistics have been developed to quantify the geometry
and topology of large-scale structure.  The topological genus of
isodensity contours characterizes the connectivity of large-scale
structure \citep{Gott:1987ab}. Measurements of the genus are consistent
with random phase initial conditions (as predicted by inflation) on
large scales \citep{Gott:1989ab}, with departures from Gaussianity on
smaller scales where nonlinear gravitational evolution and biasing of
galaxies are evident \citep{Vogeley:1994cd, Gott:2006ab}. Similar 
techniques are used to check on the Gaussianity of the CMB anisotropy
\citep{Park:2001ab, Wu:2001ab, Komatsu:2003ab}, as well as identify
foreground emission signals in CMB anisotropy data \citep{Park:2002ab}.

The void probability function, which characterizes the frequency
of completely empty regions of space \citep{White:1979ab}, has been
estimated from galaxy redshift surveys \citep{Maurogordato:1987ab,
Hoyle:2004ab}. Catalogs of voids have been constructed with objective
void finding algorithms \citep{El-Ad:1996ab, Hoyle:2002cd}.

Early investigations of the pattern of galaxy clustering dating back
to \citet{Charlier:1925ab} suggested a clustering hierarchy. 
The fractal model of clustering introduced by 
\citet[][and references therein]{Mandelbrot:1982ab} further motivated
investigation of the possibility of scale-invariant clustering of
galaxies. Results of such analyses of galaxy survey data were
controversial [compare, e.g., \citet{SylosLabini:1998ab} with 
\citet{Hatton:1999ab} and \citet{Martinez:2001ab} and references therein]. 
While fractal behavior is seen on small scales, there is fairly strong 
evidence for an approach to homogeneity in galaxy redshift and 
photometric surveys on very large scales. Thus, a simple scale-invariant 
fractal description seems to be ruled out. A multi-fractal description 
of clustering continues to provide a useful complement to other 
statistical descriptors \citep{Jones:2005ab}. Consideration of modified 
forms of the fractal picture are of interest for providing slight
non-Gaussianity on very large scales that might be needed to 
explain the very largest structures in the Universe.

Anisotropy of galaxy clustering in redshift space results from bulk
flows on large scales that amplify clustering along the line of sight
to the observer and from motions of galaxies in virialized systems
such as clusters that elongate those structures along the line of
sight \citep{Kaiser:1987ab}. \citet{Hamilton:1998ab} provides
an extensive review and \citet{Tinker:2006ab} describe recent
methods for estimating cosmological parameters from redshift-space
distortions of the correlation function or power spectrum.

The dependence of clustering statistics on properties of galaxies
provides important clues to their history of
formation and reflects the complex relationship
between the distributions of mass and luminous matter. The amplitude
of galaxy clustering is seen to vary with galaxy morphology
\citep[e.g.,][]{Davis:1976ab, Guzzo:1997ab}
and with luminosity \citep[e.g.,][]{Hamilton:1988ab, Park:1994ab}. In
recent analyses of the SDSS and 2dFGRS, these and similar trends with
color, surface brightness, and spectral type are seen
\citep{Norberg:2002ab, Zehavi:2005ab}.

Spectroscopy obtained with 8-10 m class telescopes has recently made it
possible to accurately study structure in the galaxy distribution at
higher redshift \citep{Coil:2004ab, Adelberger:2005ab, LeFevre:2005ab}.

\section{Measuring Cosmological Parameters}
\label{parameters}

\subsection{The Case for a Flat, Accelerating Universe}

As mentioned in Sec.\ \ref{dark}, observations of Type Ia supernovae
(SNeIa) provide strong evidence that the expansion of the Universe is
accelerating. Type Ia supernovae have the useful property that their
peak intrinsic luminosities are correlated with how fast they dim,
which allows them to be turned into standard candles. At redshifts
approaching unity, observations indicate that they are dimmer (and so
farther away) than would be predicted in an unaccelerating Universe
\citep{Riess:1998ab, Perlmutter:1999ab}. In the context of general
relativity this acceleration is attributed to dark energy that varies
slowly with time and space, if at all.  A mass-energy component that
maintains constant (or nearly constant) density has negative pressure.
Because pressure contributes to the active gravitational mass density,
negative pressure, if large enough, can overwhelm the attraction caused by
the usual (including dark) matter mass density and result in
accelerated expansion. For a careful review of the early supernova
tests see \citet{Leibundgut:2001ab}. For discussions of the
cosmological implications of this test see \citet{Peebles:2003ab} and
\citet{Perivolaropoulos:2006ab}. Current supernova data show that
models with vanishing cosmological constant are more than four
standard deviations away from the best fit.

The supernova test assumes general relativity and probes the geometry
of spacetime. The result is confirmed by a test using the CMB
anisotropy that must in addition assume the CDM model for structure
formation discussed in Sec.\ \ref{dark} (see Sec.\ \ref{galformation}
for apparent problems with this model). As discussed in Sec.\
\ref{anisotropy}, CMB anisotropy data on the position of the first
peak in the angular power spectrum are consistent with the curvature of
spatial hypersurfaces being small. Many independent lines of evidence
indicate that the mass density of nonrelativistic matter (CDM and
baryons) --- a number also based on the CDM structure formation model
--- is about 25 or 30 \% of the critical Einstein-de Sitter density
(see Secs.\ \ref{dark} and \ref{constraints}). Because the contemporary 
mass density of radiation and other relativistic matter is small, 
a cosmological constant or dark energy must contribute 70 or 75 \% of the
current mass budget of the Universe.  For reviews of the CMB data
constraints see \citet{Peebles:2003ab}, \citet{Copeland:2006ab}, and 
\citet{Spergel:2006ab}.

\subsection{Observational Constraints on Cosmological Parameters}
\label{constraints}

The model suggested by the SNeIa and CMB data, spatially flat and with
contemporary mass-energy budget split between a cosmological constant
or dark energy ($\sim$ 70 \%), dark matter ($\sim$ 25 \%), and
baryonic matter ($\sim$ 5 \%), is broadly consistent with the results
of a large number of other cosmological tests. In this
subsection we present a very brief discussion of some of these tests and
the constraints they impose on the parameters of this ``standard"
cosmological model.  Two nice reviews of the cosmological tests
are Sec.\ 13 of \citet{Peebles:1993ab} and 
\citet{Sandage:1995ab}. \citet{Hogg:1999ab} provides a concise
summary of various geometrical measures used in these tests. Section IV 
of \citet{Peebles:2003ab} reviews more recent developments and
observational constraints. Here we summarize some of these as well as
the significant progress of the last four years. Numerical values for
cosmological parameters are listed in \citet{Lahav:2006ab},
although in some cases there is still significant ongoing debate.

There have been many --- around 500 --- measurements of the Hubble 
constant $H_0$, \citep{Huchra:2006ab}, the current expansion rate.
Since there is debate about the error estimates of some 
of these measurements, a median statistics meta-analysis estimate of
$H_0$ is probably the most robust estimate \citep{Gott:2001ab}.
At two standard deviations this gives $H_0 = 100 h$ km s$^{-1}$
Mpc$^{-1}$ $= 68 \pm 7$  km s$^{-1}$ Mpc$^{-1}$ $= (14 \pm 1\ {\rm 
Gyr})^{-1}$ \citep{Chen:2003ab}, where the first equation 
defines $h$. It is significant that this result agrees with the 
estimate from the HST Key Project \citep{Freedman:2001ab}, the HST 
estimate of Sandage and collaborators \citep{Sandage:2006ab}, and 
the WMAP three-year data estimate (which assumes the CDM structure 
formation model) \citep{Spergel:2006ab}.

A measurement of the redshift dependence of the Hubble parameter
can be used to constrain cosmological parameters \citep{Jimenez:2002ab,
Simon:2005ab}. For applications of this test using preliminary
data see \citet{Samushia:2006ab} and \citet{Sen:2007ab}.

Expansion time tests are reviewed in 
\citet[Sec.\ IV.B.3]{Peebles:2003ab}. A recent development is the WMAP
CMB anisotropy data estimate of the age of the Universe, $t_0
= 13.7 \pm 0.3$ Gyr at two standard deviations \citep{Spergel:2006ab},
which assumes the CDM structure formation model. This WMAP $t_0$
estimate is consistent with $t_0$ estimated from globular cluster
observations \citep{Krauss:2003ab, Gratton:2003ab, Imbriani:2004ab}
and from white dwarf star measurements \citep{Hansen:2004ab}. The
above values of $H_0$ and $t_0$ are consistent with 
a spatially-flat, dark energy dominated Universe.

As discussed in Sec.\ \ref{nucleosynthesis}, \citet{Peebles:2003ab}, 
 \citet{Field:2006ab}, and \citet{Steigman:2006ab}, $^4$He and $^7$Li 
abundance measurements favor a higher baryon density than the D abundance
measurements and the WMAP CMB anisotropy data. (This difference 
is under active debate.) However, it is 
remarkable that high-redshift $(z \sim 10^3)$ CMB data and 
low-redshift ($z \lsim$ few) abundance measurements indicate a 
very similar baryon density. A summary range of the baryonic 
density parameter from nucleosynthesis
is $\Omega_B = (0.0205 \pm 0.0035) h^{-2}$ at two 
standard deviations \citep{Field:2006ab}.

As mentioned above, Type Ia supernovae apparent magnitudes as a
function of redshift may be used to constrain the cosmological
model. See \citet[Sec.\ IV.B.4]{Peebles:2003ab} for a summary of
this test. \citet{Riess:1998ab} and \citet{Perlmutter:1999ab} provided
initial constraints on a cosmological constant from this test, and
\citet{Podariu:2000ab} and \citet{Waga:2000ab} generalize the method
to constrain scalar field dark energy. Developments may be traced back
from \citet{Wang:2005ab}, \citet{Clocchiatti:2006ab},
\citet{Astier:2006ab}, \citet{Riess:2007ab}, \citet{Nesseris:2005ab},
\citet{Jassal:2006ab}, and \citet{Barger:2007ab}. Proposed satellite
experiments are under active discussion and should result in tight
constraints on dark energy and its evolution. See
\citet{Podariu:2001cd}, \citet{Perlmutter:2006ab}, and
\citet{Refregier:2006ab} for developments in this area.

The angular size of objects (e.g., quasars, compact radio sources,
radio galaxies) as a function of redshift provides another
cosmological test. These observations are not as numerous as the
supernovae, so this test is much less constraining, but the results
are consistent with those from the SNIa apparent magnitude
test. Developments may be traced back through \citet{Chen:2003cd} and
\citet{Podariu:2003ab}. \citet{Daly:2006ab} describe a way of
combining the apparent magnitude and angular size data to more tightly
constrain cosmological parameters.

``Strong'' gravitational lensing, by a foreground galaxy or cluster of
galaxies, produces multiple images of a background radio source.  The
statistics of strong lensing may be used to constrain the cosmological
model. \citet{Fukugita:1990ab} and \citet{Turner:1990ab} note that for
low nonrelativistic matter density the predicted lensing rate is
significantly larger in a cosmological constant dominated
spatially-flat model than in an open model. The scalar field dark
energy case is discussed in \citet{Ratra:1992ef} and lies between
these two limits. For reviews of the test see \citet[Sec.\
IV.B.6]{Peebles:2003ab} and \citet{Kochanek:2006cd}. Recent
developments may be traced back from \citet{Fedeli:2006ab}.
Cosmological constraints from the CLASS gravitational lens statistics
data are discussed in \citet{Chae:2002ab, Chae:2004ab}, and
\citet{Alcaniz:2005ab}. These constraints are consistent with those
derived from the supernova apparent magnitude data, but are not as
restrictive.

Galaxy motions respond to fluctuations in the gravitational potential,
thus peculiar velocities of galaxies may be used to estimate the 
nonrelativistic matter density parameter $\Omega_{\rm M}$ [as discussed 
in Secs.\ \ref{dark} and \ref{peculiar} above and in
\citet{Peebles:1999ab} and \citet[Sec.\ IV.B.7]{Peebles:2003ab}] 
by comparing the pattern of flows with maps of the galaxy distribution. 
Note that peculiar velocities are not sensitive to a homogeneously
distributed mass-energy component. For a summary of recent results from the 
literature see \citet{Pike:2005ab}. Measurements of the anisotropy 
of the redshift-space galaxy distribution that is produced by peculiar 
velocities also yield estimates of the matter density \citep{Kaiser:1987ab, 
Hamilton:1998ab}. Most methods measure this anisotropy in the galaxy 
autocorrelation or power spectrum \citep[see, e.g.,][]{Tinker:2006ab}.  
Recent analyses include \citet{Hawkins:2003ab} and \citet{daAngela:2005ab} 
from the 2dFGRS and 2QZ surveys. Also of interest are clustering analyses 
of the SDSS that explicitly take into account this redshift-space anisotropy
either by using the predicted distortions when constructing eigenmodes
\citep{Pope:2004ab} or by constructing modes that are sensitive to
radial vs.\ angular fluctuations \citep{Tegmark:2004ab}. 

A median statistics analysis of density estimates from peculiar
velocity measurements and a variety of other data indicates that the
nonrelativistic matter density parameter lies in the range $0.2 \lsim
\Omega_0 \lsim 0.35$ at two standard deviations
\citep{Chen:2003ef}. This is consistent with estimates based on other
data, e.g., the WMAP CMB data result in a very similar range
\citep{Spergel:2006ab}.

``Weak'' gravitational lensing (which mildly distorts the images of
background objects), in combination with other data, should soon
provide tight constraints on the nonrelativistic matter density
parameter. For reviews of weak lensing see \citet{Refregier:2003ab},
\citet{Schneider:2006cd}, and \citet{Munshi:2006ab}.  See
\citet{Schimd:2006ab}, \citet{Hetterscheidt:2006ab},
\citet{Kitching:2006ab} for recent developments. Weak gravitational
lensing also provides evidence for dark matter 
\citep[see, e.g.,][]{Clowe:2006ab, Massey:2007ab}.

Rich clusters of galaxies are thought to have originated from volumes
large enough to have fairly sampled both the baryons and the dark
matter. In conjunction with the nucleosynthesis estimate of the
baryonic mass density parameter, the rich cluster estimate of the
ratio of baryonic and nonrelativistic (including baryonic) matter ---
the cluster baryon fraction --- provides an estimate of the
nonrelativistic matter density parameter \citep{White:1993ab,
Fukugita:1998ab}.  Estimates of $\Omega_{\rm M}$ from this test are in
the range listed above. A promising method for measuring the cluster
baryonic gas mass fraction uses the Sunyaev-Zel'dovich effect
\citep{Carlstrom:2002ab}.

An extension of this cluster test makes use of measurements of the
rich cluster baryon mass fraction as a function of redshift. For
relaxed rich clusters (not those in the process of collapsing) the
baryon fraction should be independent of redshift. The cluster baryon
fraction depends on the angular diameter distance
\citep{Sasaki:1996ab, Pen:1997ab}, so the correct cosmological model
places clusters at the right angular diameter distances to ensure that
the cluster baryon mass fraction is independent of redshift. This test
provides a fairly restrictive constraint on $\Omega_{\rm M}$,
consistent with the range above; developments may be traced back
through \citet{Allen:2004ab}, \citet{Chen:2004ab},
\citet{Kravtsov:2005ab}, and \citet{Chang:2006ab}.  When combined with
complementary cosmological data, especially the restrictive SNIa data,
the cluster baryon mass fraction versus redshift data provide tight
constraints on the cosmological model, favoring a cosmological
constant but not yet ruling out slowly varying dark energy
\citep{Rapetti:2005ab, Alcaniz:2005cd, Wilson:2006ab}.

The number density of rich clusters of galaxies as a function of
cluster mass, both at the present epoch and as a function of redshift,
constrains the amplitude of mass fluctuations and the nonrelativistic
matter density parameter \citep[see Sec.\ IV.B.9 of][and references
therein]{Peebles:2003ab}. Current cluster data favor a matter density
parameter in the range discussed above \citep{Rosati:2002ab,
Voit:2005ab, Younger:2005ab, Borgani:2006ab}.

The rate at which large-scale structure forms could eventually provide
another direct test of the cosmological model. The cosmological
constant model is discussed in \citet{Peebles:1984ab} and some 
of the more recent textbooks listed below. The scalar field dark energy
model is not as tractable; developments may be traced from
\citet{Mainini:2003ab}, \citet{Mota:2004ab}, and \citet{Maio:2006ab}.

Measurements of CMB temperature and polarization anisotropies [see
Sec.\ \ref{anisotropy} above and Sec.\ IV.B.11 of
\citet{Peebles:2003ab}] provide some of the strongest constraints on
several cosmological model parameters. These constraints depend on the
assumed structure formation model. Current constraints are usually
based on the CDM model (or some variant of it). As discussed in
Sec.\ \ref{anisotropy}, the three-year WMAP data \citep{Hinshaw:2006ab}
provide state-of-the-art constraints \citep{Spergel:2006ab}.

Data on the large-scale power spectrum (or correlation function) of 
galaxies complement the CMB measurements by connecting the 
inhomogeneities observed at redshift $z \sim 10^3$ in the CMB to 
fluctuations in galaxy density close to $z=0$, and by relating 
fluctuations in gravitating matter to fluctuations in luminous matter 
(which is an additional complication). For a recent discussion of 
the galaxy power spectrum see \citet{Percival:2006cd}, from
which earlier developments may be traced. It is a remarkable success 
of the current cosmological model that it succeeds in providing a 
reasonable fit to both sets of data. The combination of WMAP data with 
clustering measurements from SDSS or the 2dFGRS reduces several of 
the parameter uncertainties. For recent examples of such analyses
see \citet{Tegmark:2004cd} and \citet{Doran:2007ab}. 

The peak of the galaxy power spectrum reflects the Hubble length at
matter-radiation equality and so constrains $\Omega_{\rm M} h$. The
overall shape of the spectrum is sensitive to the densities of the
different matter components (e.g., neutrinos would cause damping on
small scales) and the density of dark energy. The same physics that
leads to acoustic peaks in the CMB anisotropy causes oscillations in
the galaxy power spectrum --- or a single peak in the correlation
function.  \citet{Eisenstein:2005ab} report a three standard
deviation detection of this ``baryon acoustic oscillation'' peak at
$\sim 100h^{-1}$ Mpc in the correlation function of luminous red
galaxies (LRG's) measured in the SDSS. The resulting measurement of
$\Omega_{\rm M}$ is independent of and consistent with other low
redshift measurements and with the high redshift WMAP result. This is
remarkable given the widely different redshifts probed (LRG's probe
$z=0.35$) and notable because possible systematics are different. For
discussions of the efficacy of future measurements of the baryon
acoustic oscillation peak to constrain dark energy see
\citet{Wang:2006ab}, \citet{McDonald:2006ab}, and
\citet{Doran:2006cd}.  For constraints from a joint analysis of these
data with supernovae and CMB anisotropy data see \citet{Wang:2006cd}.

\citet{Tegmark:2006ab} include a nice description of how
the large-scale galaxy power spectrum provides independent measurement
of $\Omega_{\rm M}$ and $\Omega_{\rm B}$, which breaks several parameter
degeneracies and thereby decreases uncertainties on $\Omega_{\rm M}, h$ and
$t_0$.  A combined WMAP+SDSS analysis reduces uncertainties on the
matter density, neutrino density, and tensor-to-scalar ratio by roughly
a factor of two. See \citet{Sanchez:2006ab} for an analysis of 
the 2dFGRS large-scale structure data in conjunction with CMB measurements.

Measurements of the clustering of Lyman-$\alpha$ forest clouds
complement larger-scale constraints, such as those from the CMB and
large-scale structure, by probing the power spectrum of fluctuations
on smaller scales \citep{McDonald:2005ab}. Combining observations of
3000 SDSS Lyman-$\alpha$ forest cloud spectra with other data,
\citet{Seljak:2006ab} constrain possible variation with scale of the 
spectral index of the primordial power spectrum and find that 
Lyman-$\alpha$ cloud clustering may indicate a slightly higher power 
spectrum normalization, $\sigma_8$ (the fractional mass density 
inhomogeneity smoothed over $8h^{-1}$ Mpc), than do the WMAP data 
alone, or the WMAP data combined with large-scale structure measurements.

The presence of dark energy or non-zero spatial curvature causes time
evolution of gravitational potentials as CMB photons traverse the
Universe from their ``emission'' at $z\sim 10^3$ to today. The resulting 
net redshifts or blueshifts of photons cause extra CMB anisotropy, known as 
the Integrated Sachs-Wolfe (ISW) contribution.  This contribution
has been detected by cross-correlation of CMB anisotropy and
large-scale structure data. The resulting constraints on dark energy are
consistent with the model discussed above 
\citep[and references cited therein]{Boughn:2005ab, Gaztanaga:2006ab}. 
In principle, measurements of the ISW effect at different redshifts 
can constrain the dark energy model.  \citet{Pogosian:2006ab} 
discusses recent developments and the
potential of future ISW measurements.

\subsection{Cosmic Complementarity: Combining Measurements}
\label{complementarity}

The plethora of observational constraints on cosmological parameters
has spawned interest in statistical methods for combining these
constraints. \citet{Lewis:2002ab}, \citet{Verde:2003ab}, and
\citet{Tegmark:2004cd} discuss statistical methods employed in some of
the recent analyses described above.  Use of such advanced statistical
techniques is important because of the growing number of parameters in
current models and possible degeneracies between them in fitting the
observational data. Developments may be traced back through
\citet{Alam:2007ab}, \citet{Zhang:2007ab}, \citet{Zhao:2007ab}, 
\citet{Davis:2007ab}, \citet{Wright:2007ab}, and \citet{Kurek:2007ab}.

To describe large-scale features of the Universe (including CMB
anisotropy measured by WMAP and some smaller-angular-scale experiments,
large-scale structure in the galaxy distribution, and the SNIa 
luminosity-distance relation) the simplest version of the 
``power-law-spectrum spatially-flat $\Lambda$CDM model'' requires 
fitting no fewer than six parameters \citep{Spergel:2006ab}: 
nonrelativistic matter density parameter $\Omega_{\rm M}$, 
baryon density parameter $\Omega_{\rm B}$, 
Hubble constant $H_0$, amplitude of fluctuations $\sigma_8$, 
optical depth to reionization $\tau$, and scalar perturbation index 
$n$. This model assumes that the primordial fluctuations are 
Gaussian random phase and adiabatic.  As suggested by
its name, this model further assumes that the primordial fluctuation
spectrum is a power law (running power-spectral index independent 
of scale $dn/d\ln k =0$), the Universe
is flat ($\Omega_{\rm K}=0$), the bulk of the matter density is 
CDM ($\Omega_{\rm CDM} = \Omega_{\rm M} - \Omega_{\rm B}$) with no
contribution from hot dark matter (neutrino density $\Omega_{\nu}=0$),
and that dark energy in the form of a cosmological constant 
comprises the balance of the mass-energy density ($\Omega_{\Lambda}
= 1 - \Omega_{\rm M}$). Of course, constraints on this model
assume the validity of the CDM structure formation model.

Combinations of observations provide improved parameter constraints,
typically by breaking parameter degeneracies.  For example, the
constraints from WMAP data alone are relatively weak for $H_0, 
\Omega_{\Lambda}$ and $\Omega_{\rm K}$. Other measurements such as from
SNeIa or galaxy clusters are needed to break the degeneracy between
$\Omega_{\rm K}$ and $\Omega_{\Lambda}$, which lies approximately along
$\Omega_{\rm K} \approx - 0.3 + 0.4 \Omega_{\Lambda}$. The degeneracy 
between $\Omega_{\rm M}$ and $\sigma_8$ is broken by including weak 
lensing and cluster measurements.  The degeneracy between $\Omega_{\rm M}$ 
and $H_0$ can be removed, of course, by including a constraint on $H_0$. 
As a result, including $H_0$ data restricts the geometry to be very 
close to flat. A caveat regarding this last conclusion is that it 
assumes that the dark energy density does not evolve.

CDM-model-dependent clustering limits on baryon density 
\citep[$\Omega_{\rm B}=(0.0222\pm 0.0014) h^{-2}$ from WMAP and SDSS data 
combined at 95 \% confidence,][]{Tegmark:2006ab} are now better 
than those from light element abundance data (because of the tension 
between the $^4$He and $^7$Li data and the D data).  It is important 
that the galaxy observations complement the CMB data in such a way as 
to lessen reliance on the assumptions stated above for the ``power-law 
flat $\Lambda$CDM model.''  If the SDSS LRG $P(k)$ measurement is 
combined with WMAP data, 
then several of the prior assumptions used in the WMAP-alone analysis 
($\Omega_{\rm K}=0, \Omega_{\nu}=0$, no running of the spectral index 
$n$ of scalar fluctuations, no inflationary gravity waves, no dark 
energy temporal evolution) are not important.  A major reason for this 
is the sensitivity of the SDSS LRG $P(k)$ to the baryon acoustic scale, 
which sets a ``standard ruler'' at low redshift.

The SNeIa observations are a powerful complement to CMB anisotropy 
measurements because the degeneracy in $\Omega_{\rm M}$ versus 
$\Omega_{\Lambda}$ for SNIa measurements is almost orthogonal to that 
of the CMB. Without any assumption about the value of the Hubble constant 
but assuming that the dark energy does not evolve, combining SNIa and 
CMB anisotropy data clearly favors nearly flat cosmologies. On the other 
hand, assuming the Universe is spatially flat, combined SNIa and cluster 
baryon fraction data favors dark energy that does not evolve --- a 
cosmological constant --- see \citet{Rapetti:2005ab}, \citet{Alcaniz:2005cd}, 
and \citet{Wilson:2006ab}. 

The bottom line is that statistical analyses of these complementary
observations strongly support the flat $\Lambda$CDM cosmological
model. It is remarkable that many of the key parameters are now known
to better than 10\%. However, several weaknesses remain, as discussed
in the following, and final, section of this review. Time and lots of hard
work will tell if these weaknesses are simply details to be cleaned
up, or if they reveal genuine failings of the model, the pursuit of
which will lead to a deeper understanding of physics and/or astronomy.
It is worth recalling that, at the beginning of the previous century
just before Einstein's burst of 1905 papers, it was thought by most
physicists that classical physics fit the data pretty well.

\section{Open Questions and Missing Links}
\label{questions}

We conclude this review by emphasizing that cosmology is by no means
``solved.'' Here we list some outstanding questions, which we do not
prioritize, although the first two questions are certainly paramount
(What is most of the Universe made of?). It may interest the reader to
compare this discussion of outstanding problems in cosmology to those
discussed in 1996 \citep{Turok:1997ab}. Recent discussions of key
questions, with regard to funding for answering such questions, may be
found in reports of the \citet{NationalAcademy:2001ab,
NationalAcademy:2003ab}.

\subsection{What is ``Dark Energy"?}

As discussed in Secs.\ \ref{dark} and \ref{parameters}, there is
strong evidence that the dominant component of mass-energy is in the
form of something like Einstein's cosmological constant. In detail,
does the dark energy vary with space or time?  Data so far are
consistent with a cosmological constant with no spatial or temporal
evolution, but the constraints do not strongly exclude other
possibilities.  This uncertainty is complemented by the relatively
weak direct evidence for a spatially-flat universe; as
\citet{Wright:2006ab}, \citet{Tegmark:2006ab}, and \citet{Wang:2007ab}
point out, it is incorrect to assume $\Omega_{\rm K}=0$ when
constraining the dark energy time dependence, because observational
evidence for spatial flatness assumes that the dark energy does not
evolve.

More precisely, dark energy is often described by the XCDM
parameterization, where it is assumed to be a fluid with pressure
$p_{\rm X} = \omega_{\rm X} \rho_{\rm X}$, where $\rho_{\rm X}$ is the
energy density and $\omega_{\rm X}$ is time-independent and negative
but not necessarily $-1$ as in the $\Lambda$CDM model. This is an
inaccurate parameterization of dark energy; see \citet{Ratra:1991ab}
for a discussion of the scalar field case. In addition, dark energy
and dark matter are coupled in some models now under discussion, so
this also needs to be accounted for when comparing data and models;
see \citet{Amendola:2007ab}, \citet{Bonometto:2007ab},
\citet{Guo:2007ab}, and \citet{Balbi:2007ab} for recent discussions.

On the astronomy side, the evidence is not iron-clad; for example,
inference of the presence of dark energy from CMB anisotropy data relies
on the CDM structure formation model and the SNIa redshift-magnitude 
results require extraordinary nearly ``standard candle''-like behavior 
of the objects. Thus, work remains to be done to measure (or reject) 
dark energy spatial or temporal variation and to shore up the 
observational methods already in use.

With tighter observational constraints on ``dark energy,'' one might
hope to be guided to a more fundamental model for this construct.
At present, dark energy (as well as dark matter) appears to be somewhat 
disconnected from the rest of physics.

\subsection{What is Dark Matter?}

Astronomical observations currently constrain most of the gravitating
matter to be cold (small primeval free-streaming velocity) and
weakly interacting. Direct detection would be more satisfying and this
probably falls to laboratory physicists to pursue. The Large Hadron
Collider (LHC) may produce evidence for the supersymmetric sector that 
provides some of the most-discussed current options for the culprit.  
As mentioned in the previous question, some models allow for coupling 
between the dark matter and dark energy. On the astronomy side, 
observations may provide further clues and, perhaps already do; there 
are suggestions of problems with ``pure" CDM from the properties of 
dwarf galaxies and galactic nuclear density profiles. Better 
understanding of the complex astrophysics that connect luminous (or, 
at least, directly detectable) matter to dark matter will improve such 
constraints.

\subsection{What are the Masses of the Neutrinos?}

In contrast to various proposed candidates for the more dominant
``cold" component of dark matter, we know that neutrinos
exist. While there are indications from underground experiments of
non-zero neutrino mass \citep{Eguchi:2003ab} and the cosmological tests
discussed above yield upper bounds on the sum of masses of all light 
neutrino species, there has yet to be a detection of the effect of 
neutrinos on structure formation.  A highly model-dependent analysis of
Lyman-$\alpha$ forest clustering \citep{Seljak:2006ab} results in an
upper bound of $\sum m_{\nu}<0.17$ eV (95 \% confidence; the sum is over light
neutrino species).

\subsection{Are Constraints on Baryon Density Consistent?}

Using the standard theory for nucleosynthesis to constrain the baryon
density from observations of light element abundances, measurements of
$^4$He and $^7$Li imply a higher baryon density than do D
measurements, see Secs.\ \ref{nucleosynthesis} and \ref{constraints}
and \citet{Field:2006ab} and \citet{Steigman:2006ab}.  Constraints on 
the baryon
density from WMAP CMB anisotropy data are consistent with that from
the D abundance measurements. It is possible that more and better data
will resolve this discrepancy. On the other hand, this might be an
indication of new physics beyond the standard model.

\subsection{When and How Was the Baryon Excess Generated?}

Matter is far more common than anti-matter. It is not yet clear
how this came to be. One much-discussed option is that grand 
unification at a relatively high temperature is responsible 
for the excess. An alternate possibility is that the matter
excess was generated at much lower temperature during the 
electroweak phase transition.

\subsection{What is the Topology of Space?}

The observational constraints we have reviewed are local; they do not
constrain the global topology of space.  On the largest observable
scales, CMB anisotropy data may be used to constrain models for the 
topology of space \citep[see, e.g.,][and references cited therein]{Key:2006ab}.
Current data do not indicate a real need for going beyond the simplest 
spatially-flat Euclidean space with trivial topology.

\subsection{What Are the Initial Seeds for Structure Formation?}

The exact nature of the primordial fluctuations is still uncertain. 
The currently-favored explanation posits an inflationary epoch
that precedes the conventional Big Bang era (see Sec. \ref{inflation}). The 
simplest inflation models produce nearly scale-invariant adiabatic
perturbations.  A key constraint on inflation models is the slope of
the primordial spectrum; WMAP data \citep{Spergel:2006ab} suggest a
deviation from the scale-invariant $n = 1$ value, but this is not yet
well measured.  At present, the most promising method for observationally
probing this early epoch is through detection of (the scale-invariant 
spectrum of) inflationary gravity waves predicted in a number of inflation 
models. Detection of these waves or their effects (e.g., measuring the 
ratio of tensor to scalar fluctuations via CMB anisotropy data), would 
constrain models for inflation; however, non-detection would not rule out
inflation because there are simple inflation models without significant
gravity waves.

Another critical area for studying the initial fluctuations regards
the possibility of non-Gaussian perturbations or isocurvature (rather
than adiabatic) perturbations. The evidence indicates that these are
sub-dominant, but that does not exclude a non-vanishing and
interesting contribution.

Some models of inflation also predict primordial magnetic field
fluctuations. These can have effects in the low-redshift Universe,
including on the CMB anisotropy. Observational detection of some of
these effects will place interesting constraints on inflation.

\subsection{Did the Early Universe Inflate and Reheat?}

Probably (although we would not be astonished if the answer
turned out to be no). 
With tighter observational constraints on the fossil
fluctuations generated by quantum mechanics during inflation one 
might hope to be guided to a more fundamental model of inflation.
At present, inflation is more of a phenomenological construct;
an observationally-consistent, more fundamental model of inflation
could guide the development of very high energy physics. This 
would be a major development. Another pressing need is to have a more 
precise model for the end of inflation, when the Universe reheats
and matter and radiation are generated. It is possible that the
matter excess is generated during this reheating transition.

\subsection{When, How, and What Were the First Structures Formed?}

Discovery of evidence for the epoch of reionization, from observations
of absorption line systems toward high-redshift quasars and the 
polarization anisotropy of the CMB, has prompted intense interest, both 
theoretical and observational, in studying formation of the first 
objects. See Sec.\ \ref{galformation} above.

\subsection{How Do Baryons Light Up Galaxies and What Is Their Connection to Mass?}

Carrying on from the previous question, the details of the process of
turning this most familiar component of mass-energy into stars and
related parts of galaxies remains poorly understood. Or so it seems
when compared with the much easier task of predicting how
collisionless dark matter clusters in a Universe dominated by dark
matter and dark energy.  Important problems include the effects of
``feedback'' from star formation and active galactic nuclei, cosmic
reionization, radiative transfer, and the effect of baryons on halo
profiles. High-resolution hydrodynamic simulations are getting better,
but even Moore's law will not help much in the very near future
\citep[see comment in][]{Gott:2006ab}. Solving these problems is
critical, not only for understanding galaxy formation, but also for
using galaxies --- the ``atoms of cosmology" --- as a probe of the
properties of dark matter and dark energy.

Clues to the relationship between mass and light and, therefore, strong
constraints on models of galaxy formation, include the detailed 
dependence of galaxy properties on environment. Outstanding puzzles
include the observation that, while galaxy morphology and luminosity
strongly vary with environment, the properties of early-type (elliptical
and S0) galaxies (particularly their colors) are remarkably insensitive 
to environment \citep{Park:2007ab}.

\subsection{How Do Galaxies and Black Holes Coevolve?}

It is now clear that nearly every sufficiently massive galaxy harbors
a supermassive black hole in its core. The masses of the central 
supermassive black holes are found to correlate strongly with properties 
of the host galaxy, including bulge velocity dispersion 
\citep{Ferrarese:2000ab, Gebhardt:2000ab}. Thus, galaxy formation and the
formation and feeding of black holes are intimately related 
\citep[see, e.g.,][]{Silk:1998ab, Kauffmann:2000ab, Begelman:2005ab}.

\subsection{Does the Gaussian, Adiabatic CDM Structure Formation 
Model Have a Real Flaw?}

This model works quite well on large scales. However, on small scales
it appears to have too much power at low redshift (excessively cuspy
halo cores, excessively large galactic central densities, and too many
low-mass satellites of massive galaxies).  Modifications of the power
spectrum to alleviate this excess small-scale power cause too little
power at high redshift and thus delay formation of clusters,
galaxies, and Lyman-$\alpha$ clouds. Definitive resolution of this issue
will require more and better observational data as well as improved
theoretical modeling. If the CDM structure formation model is found to be
inadequate, this might have significant implications for a number of
cosmological tests that assume the validity of this model.

\subsection{Is the Low Quadrupole Moment of the CMB Anisotropy a
Problem for Flat $\Lambda$CDM?}

The small amplitude of the quadrupole moment observed by COBE persists
in the WMAP observations even after many rounds of reanalysis of
possible foreground contributions \citep[see][and references cited
therein]{Park:2006cd}. Although one cannot, by definition, rule out
the possibility that it is simply a statistical fluke (with
significance of about 95 \% in flat $\Lambda$CDM), this anomaly
inspires searches for alternative models, including multiply-connected
Universes (see above).

\subsection{Are the Largest Observed Structures a Problem For 
Flat $\Lambda$CDM?}

The largest superclusters, e.g., the ``Sloan Great Wall''
\citep{Gott:2005ab}, seen in galaxy redshift surveys are not
reproduced by simulations of the concordance flat $\Lambda$CDM
cosmology \citep{Einasto:2006ab}.  Perhaps we need larger simulations
\citep[see discussion in][]{Gott:2006ab} or better understanding of
how galaxies trace mass.

\subsection{Why Do We Live Just Now?}

Last, because we see the Universe from only one place, at only one
time, we must wrestle with questions related to whether or not we (or
at least our location) is special.

\citet{Peebles:2003cd} notes the remarkable coincidences that 
we observe the Universe when (1) it has just begun making a transition 
from being dominated by matter to being dominated by dark energy, (2) 
the Milky Way is just running out of gas for forming stars and planetary 
systems, and (3) galaxies have just become useful tracers of mass.
While anthropic arguments have been put forward to answer the question
of why we appear to live at a special time in the history of the
Universe, a physically motivated answer might be more productive 
and satisfying.  Understanding of the details of structure formation, 
including conversion of baryons to stars (mentioned above), and 
constraints on possible evolution of the components of mass-energy in 
the Universe may provide clues.

Progress in cosmology is likely to come from more and higher-quality
observational and simulation data as well as from new ideas. A number 
of ground-based, space-based, and numerical experiments continue to 
collect data and new near-future particle physics, cosmology, astronomy,
and numerical experiments are eagerly anticipated. It is less 
straightforward to predict when a significant new idea might emerge.  

\acknowledgments

We are indebted to L.\ Page, J.\ Peebles, and L.\ Weaver for detailed
comments on drafts of this review. We acknowledge the advice and help
of T.\ Bolton, R.\ Cen, G.\ Horton-Smith, T.\ Kahniashvili, D.\
Lambert, I.\ Litvinyuk, L.\ Page, J.\ Peebles, G.\ Richards, M.\
Strauss, R.\ Sunyaev, and L.\ Weaver. We thank E.\ Mamikonyan for
technical support. We thank the referee, Malcolm Longair, for his
helpful suggestions on this review. B.\ R.\ acknowledges support of
DOE grant DE-FG03-99EP41093.  M.\ S.\ V.\ acknowledges support of NASA
grant NAG-12243 and NSF grant AST-0507463 and the hospitality of the
Department of Astrophysical Sciences at Princeton University during
sabbatical leave.

\pagebreak

\end{document}